# GEOTRAVEL: HARVESTING AMBIENT GEOGRAPHIC FOOTPRINTS FROM GPS TRAJECTORIES

LIEW LI CHING
GOH ONG SING

FACULTY OF INFORMATION AND COMMUNICATION TECHNOLOGY
UNIVERSITI TEKNIKAL MALAYSIA MELAKA

## ABSTRACT

The study is about harvesting point of interest from GPS trajectories. Trajectories are the paths that moving objects move by follow through space in a function of time while GPS trajectories generally are point-sequences with geographic coordinates, time stamp, speed and heading. User can get information from GPS enable device. For example, user can acquire present location, search the information around them and design driving routes to a destination and thus design travel itineraries. By sharing GPS logs among each other, people are able to find some places that attract them from other people's travel route. Analysis on the GPS logs can get the point of interest that is popular. By present the point of interest, user can choose travel place easily and the travel itineraries is plan based on the user preferences.

**TABLE OF CONTENTS**

# LIST OF TABLES

## LIST OF FIGURE

**LIST OF ABBREVIATIONS**

| | | |
|---|---|---|
| ADT | – | Android Development Tools |
| API | – | Application Programming Interface |
| APK | – | Android Application Package File |
| DFD | – | Data Flow Diagram |
| GPS | – | Global Positioning System |
| IDE | – | Integrated Development Environment |
| JDK | – | Java Development Kit |
| JDT | – | Java Development Tools |
| MB | – | Megabytes |
| OS | – | Operating System |
| POI | – | Point of Interest |
| SDK | – | Software Development Kit |
| SDT | – | Software Development Team |
| UI | – | User Interface |
| UML | – | Unified Modelling Language |
| XML | – | Extensible Markup Laguage |

## LIST OF ATTACHMENT

# CHAPTER 1

# INTRODUCTION

## 1.1 Project Background

Smartphones nowadays are integrated with advanced technology that make smartphone is not a wanted item but is a needed item. Smartphone make people life more convenient and easy, everything become simple once you own a smartphone. People can use smartphone to surf internet, pay bill, e-learning and many thing else. One of the important thing that can be done by smartphone is smartphone can act as a GPS (Global Positioning System) device. A lots of GPS navigator software are available in the market. The GPS navigator navigate the route from one place to another place. The shortcoming of the GPS navigator software is the user do not know how other people travel or what is the suggested route. People may want to know how other people travel and where is the interesting place.

Trajectory mean path of moving object that follows through space as a function of time while GPS trajectories is the point sequences with geographic coordinate which someone had travel along. GPS device is one of the examples of location-acquisition technologies. By using GPS device, people will be able to record their journey or location history with a sequence of time-stamped. From other people GPS trajectories, someone can know life interests and preferences of that user, thus facilated people to do things. Trajectories can link peopele in the phiysical world by location.

Personal GPS tracking trajectories generally are point-sequences with geographic coordinates, time stamp, speed and heading. Automatic recorded GPS travel data can not only reduce the burden of persons who participant in travel survey, but also provided detailed and accurate information in time, geo-locations and route choices. However, several data processing steps are required to explore knowledge hidden in raw GPS trajectories. One of these procedures is Activity Identification which aims at discovering activities in trajectories since travel purposes are obviously not included in GPS traces. And as the volume of historic data keep increasing, completely computer aided methods are need.

By analysis on GPS trajectory, people city travel sequence can be known. This is different from the traditional travel data. The raw GPS trajectories do not contain information about trip purpose or activity, it just include GPS coordinate and time. Earlier studies addressed this issue through a combination of manual and computer-assisted data processing steps. Nevertheless, geographic context databases provide the possibility for automatic activity identification based on GPS trajectories since each activity is uniquely defined by a set of features such as location and duration. Increasing in the usage of GPS capabilities device, positions of mobile objects can be detected.

GeoTravel in harvesting ambient geographic footprints from GPS trajectories project is about analysis GPS trajectories form various user and getting the POI (point of interest) of the users. Stay point is detected from the GPS trajectories and the result is clustered by density based clustering (OPTICST). GeoTravel application can detect the location of the user automatically and therefore can give some suggestion on interesting locations. Classical travel sequences by the users can be known by GPS trajectories. This will help people who wish to plan travelling trip by themselves.

In conclusion, the dataset of GPS trajectories will be first collected and follow by analyzed. From the analysis, we will get the top interesting location (point of interest) and travel sequence among location. From that, we can transfer the point of interest into customize itinerary which will generates multiday itineraries for the users.

### 1.2 Problem Statements

Here discuss with the problem statements that happen in the real life scenarios that proves the need of building a GeoTravel application to make the trip planning easier.

i. **Users need a system that can auto detect their location**

The users need to enter the location in order to search the interesting place in some of the system/apps. It is time wasted and complicated if the users want to search the place of interest which is nearby them.

ii. **Users need effective system that can suggest the place of interest and live event**

Most of the traveling planner application do not include live event. The live event that held in a place is important for a travelers because some of the traveler especially backpacker may want to join local event to gain more experience.

iii. **Users especially backpacker need an efficient and economic trip plan application**

Creating and efficient and economic trip plan is the most annoying job for a backpacker traveler. Although travel agency can provide some predefined itineraries, they are not tailored for each specific customer.

iv. **Users or travelers need a system that can generates multiday itineraries**

In most of the existing itineraries planning system, it only provides single day itineraries but in the real world, most of the people will travel in one place more than one day. Therefore, travelers need a system that can help them plan a multiday trip.

v. **A traveler used application could be better**

There are similar applications for travelers use is published in Play Store. However, that is still not good enough for travelers. It should be more user-friendly and cover more things about the point of interest of the place.

## 1.3 Objective

The problem stated perhaps can be solved by developing this application for travelers. Here states with the objectives of building this application.

i. **To access user location by IP address and GPS trajectories**
The system should be able to get the users/travelers location automatically so that it can suggest the place of interests to the users based on the location. This will save the time of the user on insert the location manually.

ii. **To understand GPS trajectories**
Study and understand GPS trajectories from previous travelers not only can know the life interests and preferences of the traveler, it also facilitated people to do many things. From the GPS trajectories, we can suggest the traveler the common travel plan and include live event in the place of interests list.

iii. **To build an efficient and economic trip plan application**
Analysis the data of GPS trajectories will give popular places that traveled by most travelers. Recommendation is made based on the analysis result. User can plan their trip base on the recommendation.

iv. **To build a system that can generate multiday itineraries**
In most of the existing application, it only provides single day itineraries. This application can generate multiday itineraries with no repeat of place of interest. This helps most of the users because normally a trip would not be only one day.

v. **To build a user friendly and high effective application for traveler use**
This application will focus on graphical presentation. That is, interesting icons and colourful features are introduced in the application to attract users' eyesight. Other than just plan the itineraries of the trip, this application also give recommendation on the traffic.

## 1.4 Scopes

i. **User**
The system is mainly developed for the users such as below:

- Backpacker
  They are one of the main end-users of this application which the function of the system is built especially for their usage. They can plan their multiday itineraries based on the point of interests that provided by this application. Besides that, they also can include the live events that occur at that time into their trip. This application will calculate the shortest path to use for the users.

- Local Traveler
  Local traveler is the people that wish to travel in his own place. Usually local people aim is to find the live event or restaurant in the town. This application provide point of interests include restaurant and live event for the user to choose. This application will plan the itineraries based on the user selection, therefore different user preferences will not affect other.

ii. **Platform**

- Web-based

This system is a web-based system. User can surf the website by internet browser. Every information can be look through the web site. User can plan their trip by using web system.

- Android

  In order to simplify the system, android application is built. This android application can run on all android base smart-phone. Other platform such as Mac OS is not available for this moment.

**1.5 Project significance**

GeoTravel Application is built to improve the efficiency for travelers especially backpackers to plan their multiday itineraries. The point of interests that suggest to the user is based on other people trajectories history. This application provide several features like

- Get the users location
- Get the users GPS trajectories
- Analysis the GPS trajectories from other travelers
- Suggest point of interest and live events
- Perform multiday itineraries based on the user preference
- Suggest the best route for the user
- Advice user on the traffic
- Cost estimation

GeoTravel Application used a useful method to analyze the GPS trajectories from other traveler. It will give the result of the top interesting location and common travel sequence among location. It is easy to plan your own trip with this system. User only needs to select the point of interest that wish to travel and the system will automatic give the itineraries as the result.

The most beneficial people of this system are the backpackers. By this system, travelers can choose its own packages that suit them. Although the travel agencies may provide efficient and convenient services, for some travelers or backpackers, the itineraries provided by the travel agents lack customization and cannot satisfy individual requirement. Some interested point of interest even missing in the itineraries and the package are too expensive for a backpack traveler. Therefore, this system provide an alternative way for them to plan their trips in every detail, such as picking point of interest for visiting and selecting the hotels.

## 1.6 Expected output

A better efficiency and user friendly android mobile application that provide the features that help the backpacker to plan their trip. The application will auto detect the location of the user and direct to the point of interests in that location. The point of interest is suggested based on the result from other user GPS trajectories. It also includes the live events in the point of interest that help the user which are not from that place. This application is a multiday itineraries planning instead of single day. There is a limit hour to travel in each day. In order to make the system more complete, suggestion of hotel is makes in multiday itineraries. This system is built to overcome the weakness form the other published travel use application that is available in Play Store.

## 1.7 Conclusion

The project background is discussed in this chapter, thus detect the problems faced by the current situation of the travelers. Objective and scope of this project is determined to emphasize the need to build up a system like GeoTravel application so that the problem can be overcome and improve the system to be better.

The systems will first analysis the data from the previous traveler and make a conclusion of the point of interests. User can choose their own point of interest and the system will generate itineraries based on their preference point of interest. The system not only can generate single day itineraries, it also can generate multiday itineraries.

For the next chapter, the literature review and project methodology will be discussed. It also describes more details about the differentiation between existing application and GeoTravel system. Besides that, next chapter also include the requirement to build this application, the project schedule and milestone.

# CHAPTER II

# LITERATURE REVIEW AND PROJECT METHODOLOGY

## 2.1 Introduction

This chapter explains about literature review conducted and methodology that is used to develop GeoTravel Application.

According to the Literature review is about discusses published information in a particular subject area or topic. It may just a simple summary which is a recap of the important information of the sources or combines both summary and synthesis. Synthesis is a re-organization and reshuffling, of the information. It may give a new interpretation l or combine new interpretation with old one. Besides that, synthesis might trace the intellectual progression of certain field which includes major debates. Furthermore, the literature review may evaluate the source and advise the reader on the most pertinent or relevant".

According to A. Stefanidis (2012), "the using of the location-acquisition technologies like GPS device, people are able to record their trajectories which are their location history in a sequence of time-stamped. These trajectories have facilitated people to do many things based on users' life interests and preferences, and. Therefore, by using these trajectories, we not only connect locations in the physical world but also bridge the gap between users and locations."

According to Y. Zheng, “In the real world, traveler leave their location history in a form of trajectories. Based on GPS trajectories, people can share their life experience and generic travel recommendation.

“Personal GPS tracking trajectories generally are point-sequences with geographic coordinates, time stamp, speed and heading. Automatic recorded GPS travel data can not only reduce the burden of persons who participant in travel survey, but also provided detailed and accurate information in time, geo-locations and route choices. However, several data processing steps are required to explore knowledge hidden in raw GPS trajectories. One of these procedures is Activity Identification which aims at discovering activities in trajectories since travel purposes is obviously not included in GPS traces. And as the volume of historic data keep increasing, completely computer aided methods are need”, said by L.Huang (2010).

In brief, there are many methodologies are applicable in software management project, such as Waterfall model, agile management, Spiral model and also Rational Unified Process (RUP). The size, composition, priorities, criticality and skills that are different in each project are always the factor to be concerned about while selecting a methodology. An appropriate methodology would optimize the project management process by introducing a framework that may reduce the risk level. Hence, the methodology that is used to develop GeoTravel Application will introduced in this chapter as it does play the main key in ensuring the success of this project development.

## 2.2 Facts and findings (based on topic)

A practical solution for the trip planner application is to provide an automatic itinerary planning service. The user lists a set of interested point of interests (POIs) and specifies the time and money budget. The itinerary planning service is then returns top-K trip plans satisfying the requirements. In the ideal case, the user selects one of the returned itineraries as his plan and notifies the agent. There are a lot of algorithms for automatic itinerary planning service. However, only few of the current itinerary planning can generate a ready-to-use trip plan, as they are based on various assumptions. The one that can generate a ready-to-use trip plan is not based on the GPS trajectories of the previous travelers.

With GPS device like GPS phones have become popular in modern life. People always used GPS device during travelling. Hence, a lot of GPS log data which include GPS trajectories have been accumulated both continuously and unobtrusively. There is many new generate application regarding GPS currently due to the increasing of the usage of GPS device which collect GPS data based on real world. Users are allowed to view, upload and share GPS tracks logs. Multimedia content in the application or web also attract people to use the device.

GPS trajectories contain information about the travel sequence of the users. By analysis on GPS trajectories, travel recommendation can be generic. This can be done by mining the data based on GPS trajectories. Based on the result, interesting location and classical travel sequence can be detected in a given place. The interesting location or point of interest in here include culturally place like museum, events and common frequented public area such as restaurant.

Current planning algorithm of the trip planner normally only consider a single day's trip, while in real cases, most users will schedule an *n*-day itinerary. Generating an *n*-day itinerary is more complex than generating a single day one because it is not just combining the single day itinerary to become *n*-day itinerary. This is because point of interest can only appear once in the itinerary. One of the possible solutions is to exploit the geolocation (put the nearby POIs in the same day's itinerary). Besides that, we can also rank POIs by using a priority queue base on their importance to schedule the trip.

Novel itinerary planning algorithm is used to generate itineraries that narrow the gap between the agents and travelers. This algorithm reduces the overhead of constructing a personalized itinerary for the traveler and also provides a tool for the agents to customize their service.

Besides that, smartphone tend to become necessary item in our daily life. Number of mobile application in the market keep on increasing, therefore choosing a mobile platform become an issue for developer. There are few platforms which are quite popular in the market nowadays. These platforms include iOS, Android, Lumnia, Symbian and Bada. Among all these, iOS and Android always bear to mind first than others. iOS is mobile operating system of Apple Inc. while Android is introduced by Google Inc.. Bother operating systems provide high user-friendly, well designed and also high efficiency. For this project, Android will be chosen because it is an open-source development environment. Meanwhile, its functionality and power is almost same to iOS.

### 2.2.1 Domain

- **GPS Trajectories**

  There is a branch of GPS trajectories related application available on the market or internet. By using those applications by GPS device, people can record or view their travel routes. Users can also share their travel experiences by upload their GPS trajectories to the web or application. Previously, people shared their experience in text-based article or blog. GPS trajectories applications give an alternative method that is more interactive than the previous one. This method helps users with valuable references when planning a travel itinerary.

- **Web-based system**

  Web-based system can be done by using many web languages such as PHP and java script. In this system, the language use is PHP. User can browse through the web site by any browser. Point of interest and event can be seeing through the web. User can also plan their trip by using the web based system.

- **Android**

  Android is an open platform that had become famous recently. It drives innovation in mobile communications and beyond. Android application is easy to be built because it is an open source application such as Linux community and other hardware, software and carrier partners. Due to android is an open source application, it has rapidly grow and become a famous platform to be used recently. People are able to develop any android application by various platforms since it was an open platform. Developer can published their application through open marketplace.

- **Android Application**

  The simple software that is built to be run on the Android platform such as in smartphone or tablet which provide entertainments, tools and also platform for internet browsing. Most of the applications can be found in Google Play Market and download for free or either with charge.

- **Android API Level**

  Android API Level is an integer value that uniquely identifies the framework. The API revision is offered by a version of the Android platform. The framework API consists of:

  - A core set of packages and classes
  - A set of XML elements and attributes for declaring a manifest file
  - A set of XML elements and attributes for declaring and accessing resources
  - A set of Intents
  - A set of permission that application can request, as well as permission enforcements included in the system
  - Each successive version of the Android platform can include updated to the Android application framework API that it delivers.

### 2.2.2 Existing System

There are some travel use applications that exists in the market. In this topic, 3 selected applications or algorithm used are discussed here.

i. GeoLife

In GeoLife application, GPS data are collected through a device. Those data are real data and are well managed. The result is easy understood as compared to current technologies. Visualizing GPS data over digital maps and display result make this application interactive. Besides that, this application suggests the point of interest based on the user behavior or travel history. Searching function is also available in this application. The GPS trajectories show the search result over digital maps by using spatial range or a temporal interval as query. By these, people can understand other peoples' travel sequence or live pattern from their GPS trajectories.

This application is quite functional in the GPS trajectories field but not in trip planner application. This application only gets the raw GPS trajectories and does analysis from the GPS trajectories. The result get from the analysis is then transforms into point of interests. From there we can know classic routes, popular travel routes and point of interest. The shortage of this application is users cannot customize the place that they want to go, they just can read the history of GPS trajectories from the past.

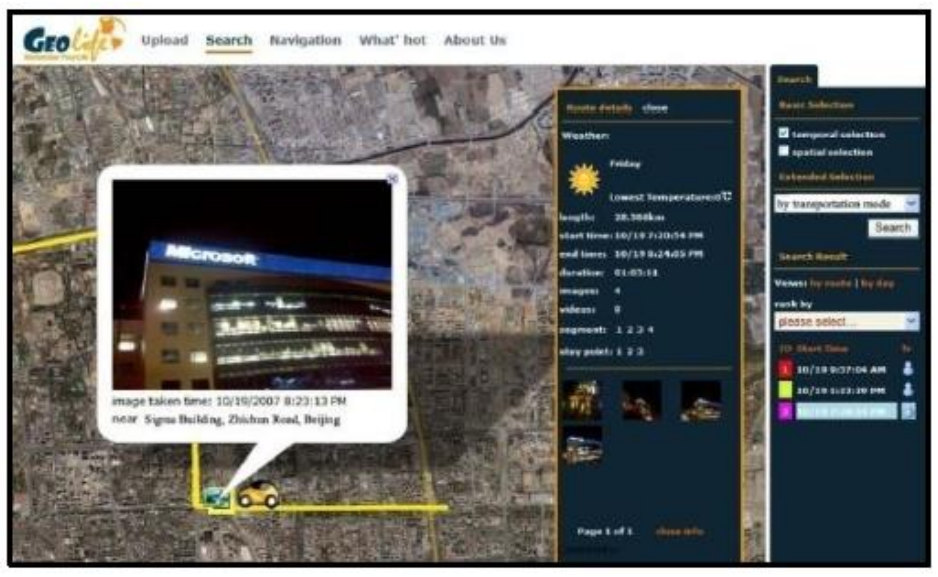

**Figure 2.1: GeoLife Application**

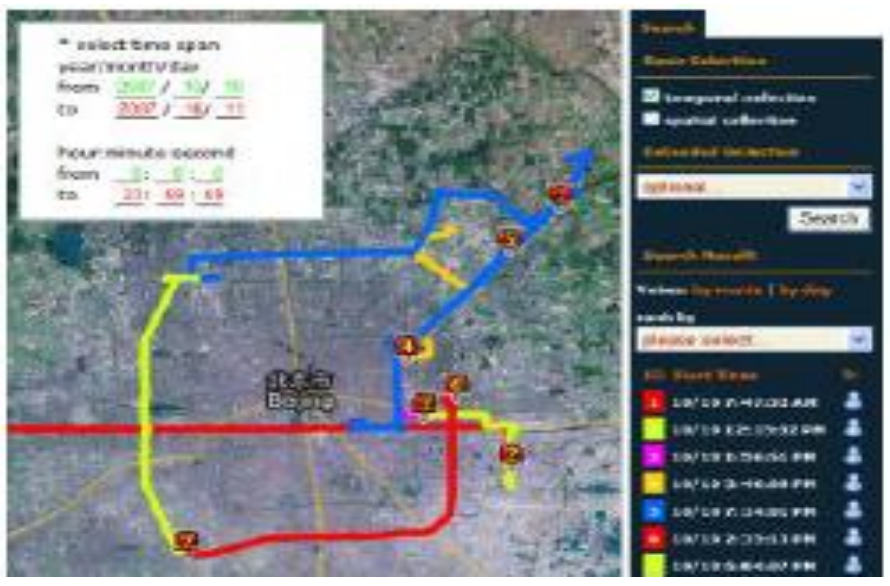

**Figure 2.2: Searching Trajectories**

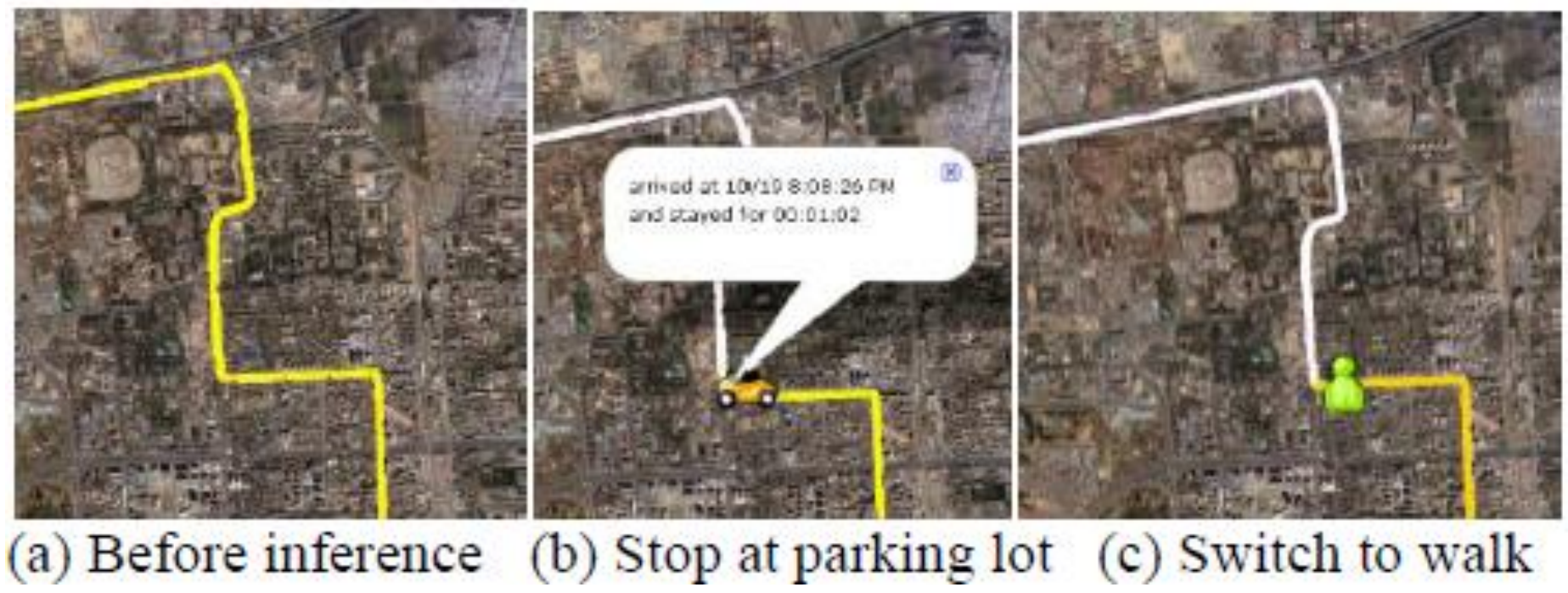

**Figure 2.3: GPS Trajectories data**

ii. Trip Planner

Tourism Malaysia had launched a travel application which is Trip Planner application. With this application, panning a memorable holiday becomes simple and interesting. These applications describe the points of interest and show the unique attractions around Malaysia. Besides that, this application also highlights special event or cultural festival in different time. By using this application, users can add items to their itinerary, save them and share with their social networks.

This application show few point of interests in Malaysia. User can choose which point of interest mainly cities to explore and add in their itinerary. After user has chosen, the information about that point of interest is shown and some nearby attraction is shown in the map. User can know more information about that point of interest via this application.

Although this application provide a lot of information about the point of interest but the point of interest of this application is lesser if compare to other application. Besides that, this application does not suggest the route and plan the itinerary for the user. This application does not help to user to plan their trip by which point of interest the user can go in one day. This application only shows the location of the point of interest on the map and the users need to find it by their own.

Although this application is official travel application, but it lack of live event that occur on certain date. Many tourists may need to want to join the live event on the city to gain more experience of the city. This application should show how to get to the certain point of interest by GPS navigator because travelers may need to know how to go from one place to another place.

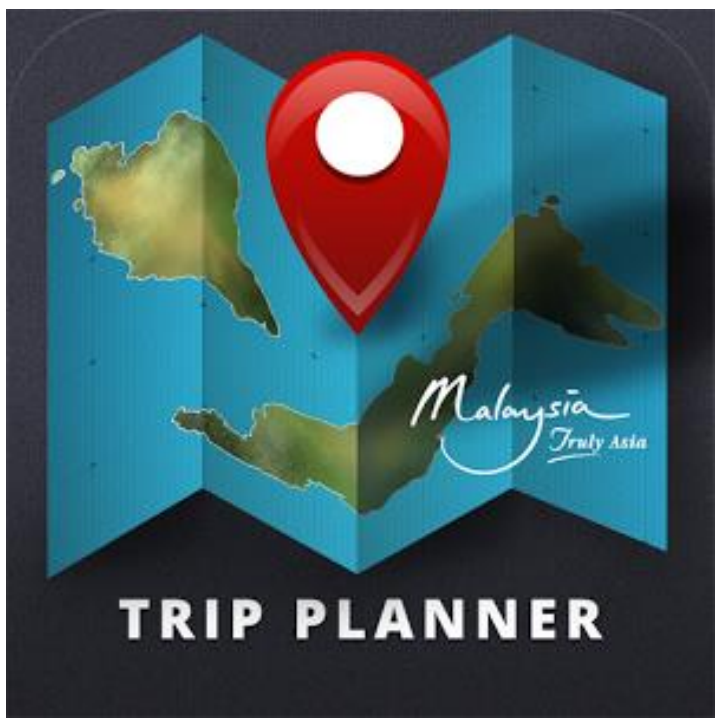

Figure 2.4: Trip Planner

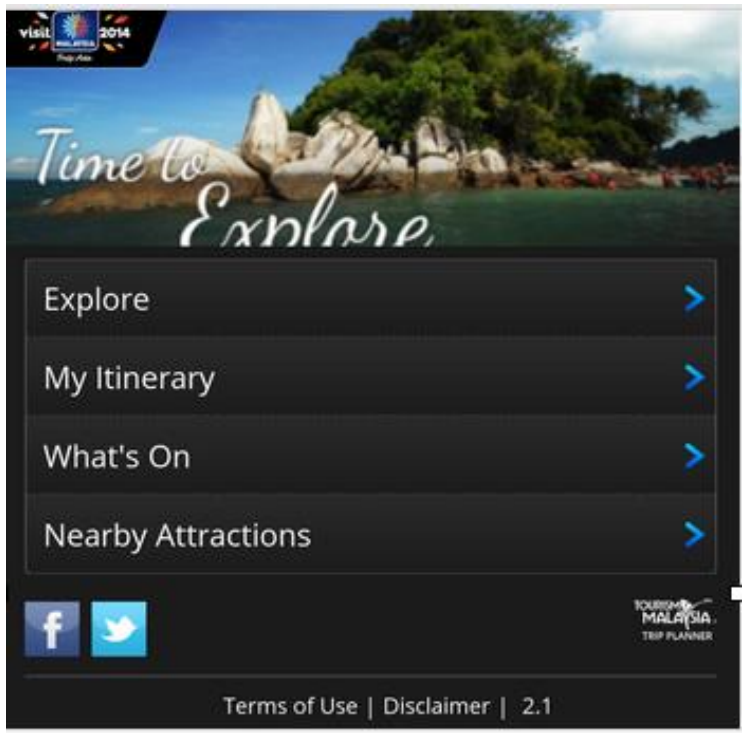

Figure 2. 5: Main Menu

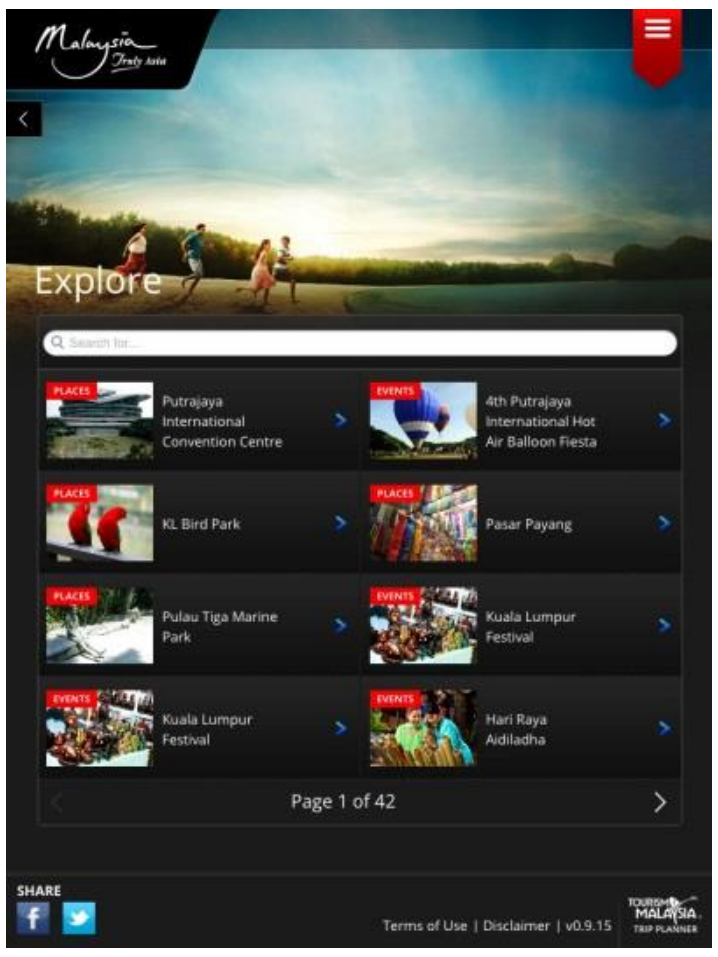

Figure 2.6: Point of interest

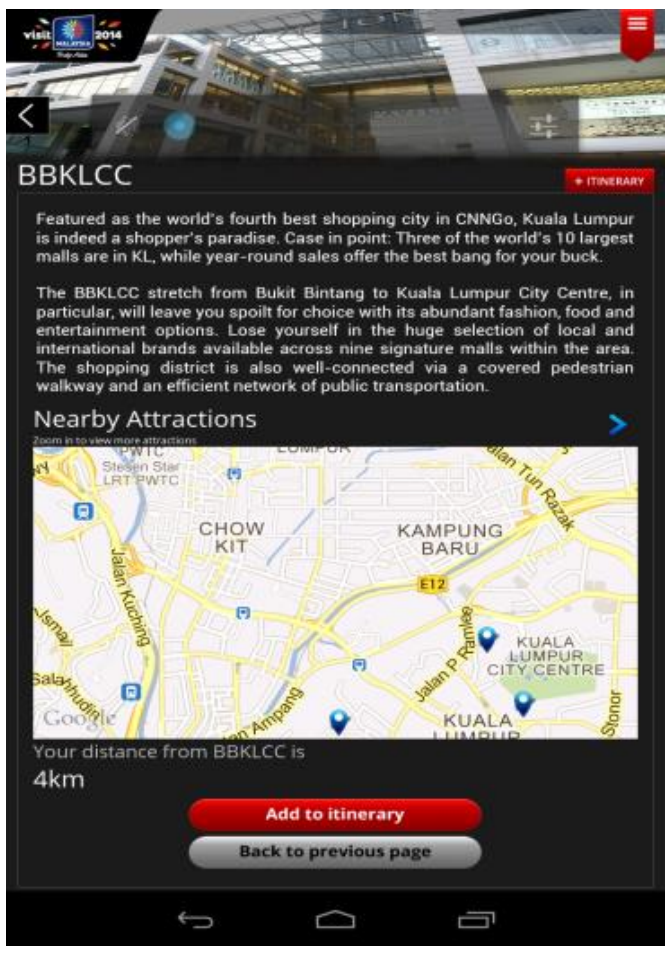

Figure 2.7: Location by map

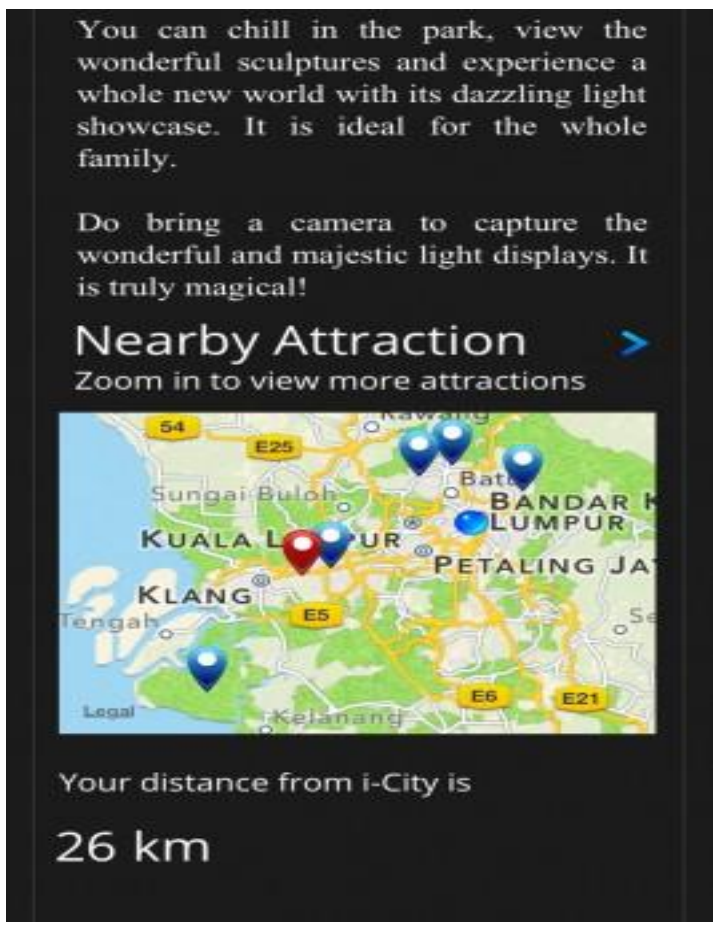

Figure 2.8: Nearby Attraction

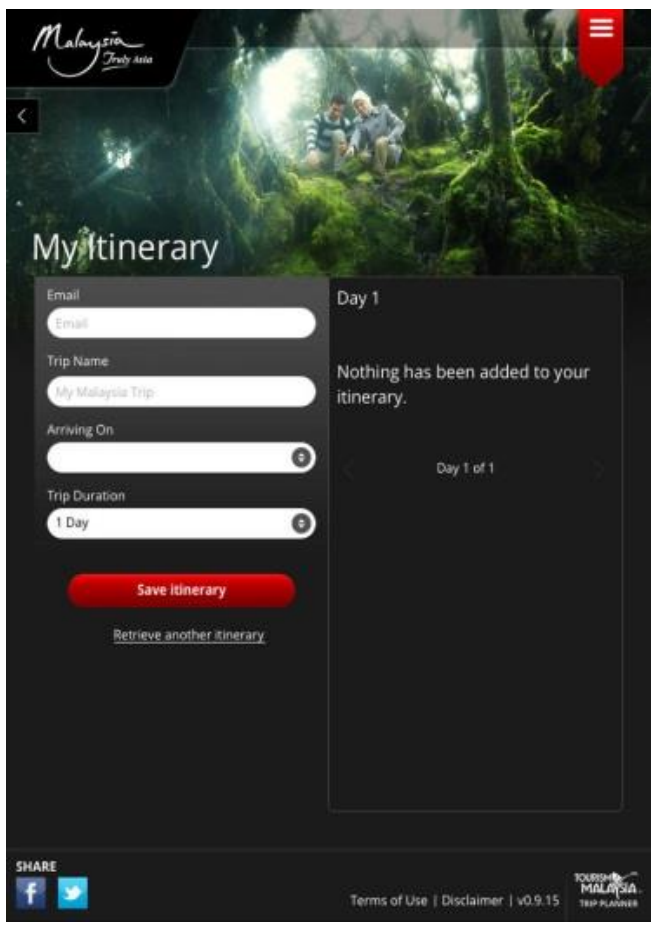

Figure 2.9: Itinerary

iii. Tripomatic

Tripomation application is an international application. It shows the point of interest along the world. This application is type of sightseeing travel guide. It Discover thousands of attractions in 400+ destinations worldwide. Users can plan trip itineraries either one day trip or multiple days by this application. All information is automatically synced to www.tripomatic.com so that users can edit on the web without using the device. Besides that, offline maps are also available.

The unique or advantage of Tripomatic's application is that the information or trip planning are sync up with a web-based itinerary-planning service. Users can know the point of interest when travel to other place or country. This application contains the information of different fields. This gives freedom to the traveler or blacker to explore while on vacation.

Users can user Tripomatic application to can make personal travel itinerary by choosing the destination and pick sights the point of interest which their wish to visit or interesting to them.

By using this application, user can:

- Find nearby point of interest or attraction
- Get the details of point of interest include contacts, business hours and admission fees
- Get personal travel itineraries plan
- Plan trip by using phone
- Edit trip by using web
- Get offline map
- Included more 40000 attractions worldwide

This application gives an alternative choice for the trip planner. The shortage of this application is it is a worldwide application therefore it only contain point of interest that from the main city. If the traveler wish to travel in small town of certain country, the information that provide by this application may not sufficient. The lacks of information of local event also make this system not a perfect application.

**Figure 2.10: Tripomatic Application**

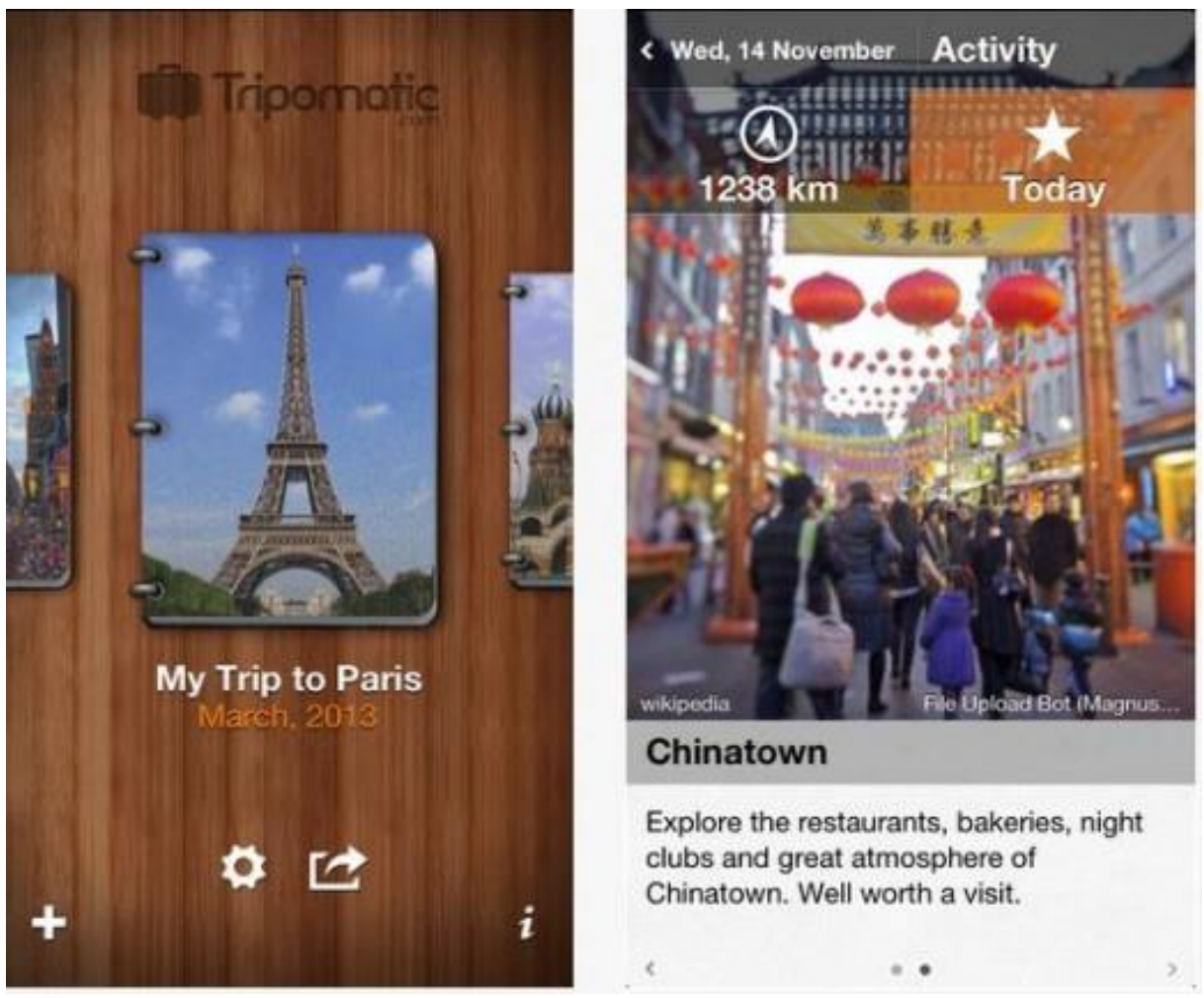

**Figure 2.11: Chose city to travel**

### 2.2.3 Technique

Online research and analysis of algorithm is done to collect the algorithm that suitable for the system. After study about few algorithms, one of the most suitable algorithms is chose to implements in the system. Analyses the weakness and strengths of the other algorithms will help to build up a better algorithm by improve it.

Online research is also done to collect information about Android application development. Study on gaining user experience in order to achieve higher degree of user friendliness and usability is done too by using other existing traveler use applications that are published on the Play Market. Analyses the weakness and strengths of the other application will help to build up a better software application product by improve it.

### 2.3 Project Methodology

Methodology is the systematic and theoretical analysis of the methods which applied to a field or study. In this project, methodology is used to gather relevant data about GPS trajectories from specified documents. The data is then analyzed and the result is discussed. Databases are compiled by using the point of interest from the analysis process. By undergo all these project will improve understanding about GPS trajectories. This project will utilize both quantitative and qualitative data collection tools, but is rooted in a qualitative epistemological position that recognizes the importance of locating the research within a particular social, cultural, and historical context. It also takes seriously the social construction of these contexts and the identities participants construct within them.

Agile methodology is used in this project development. Agile methodology is an incremental, repetitious means of managing projects particularly in the field of software development. It was built upon the foundations of the traditional waterfall sequential methods of the 1970’s. In the late 1970’s and throughout the 1980’s, Japanese companies began using phased program planning in new product development where each phase of the project is continue to be revisited throughout the project until the project reach the final phase in a short duration. This led to just-in-time manufacturing and eventually becomes agile software development

Agile software development methodology is divided into two which is XP and Serum. XP stands for extreme programming and it concentrates on the development rather than managerial aspects of software projects. XP development starts with release planning phase, followed by several iterations which concludes with user acceptance testing. This project is more to XP style of agile development where the development phase is concerned more. Refer figure 2.12 for the more detail.

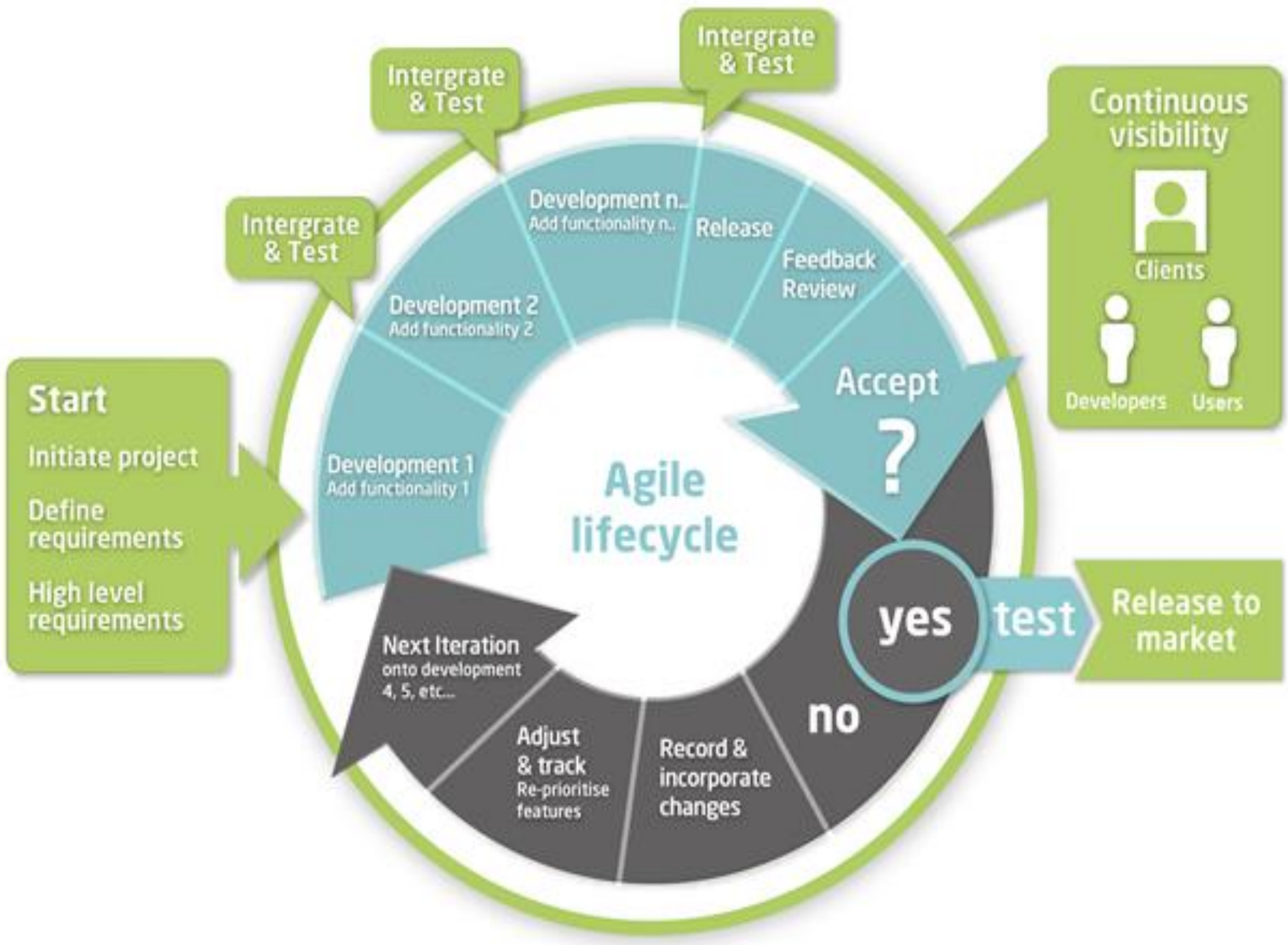

**Figure 2.12: Agile Development**

This type of methodology introduces a set of XP rules and concepts, which is integrate often, project velocity, pair programming and user story. Integrate often is also called as continuous integration. The changes in continuous integration must be integrated into the development baseline frequently. Besides, project velocity is important in every development. Project velocity measures the work along the project development. This process makes sure the product can be delivery on date by drives a proper planning and gets the schedule updated. Pair Programming is where two people working together in a single computer in production of code in order to achieve higher quality of user satisfaction. Gather user story also important because that describes the problems faced by them and need to be solved by the system being built. This story is written by user and should be about three sentences long. One of the strengths of XP development is able to figure out what the user is not clear about the criteria of the system to be built.

In the first stage of the system, users' stories which describe the need of the software are collected. The system should fulfill the user stories. After this process, the time to build and the resources needed to build the system is estimated. By using the information, a release plan is created and the development tasks are divided into iterations. The release plan defines each iteration plan in order to drives for the iteration. After test case being done by the developer, the user performs acceptance tests based on the user stories. If the users accept the system, the system is released else, the system being modified to fulfill the needs of the users. If the is bug in the system, then bug fixing is the next iterantion.

## 2.4 Project Requirements

Here introduce with the project requirement. There are software requirement and hardware requirement in this part. Software requirement is the software that needed to build the system while hardware requirement is hardware that necessary for the development of system.

### 2.4.1 Software Requirement

Refer table 2.1 for descriptions of software requirement in this project.

**Table 2.1: Software Requirement**

| Item | Description |
| --- | --- |
| Microsoft Windows 7 | Operating system from Microsoft |
| Google Chrome/Mozilla Firefox | Web Browser |
| WordPress | Website creator |
| Eclipse Kepler | IDE for Android and Java development |
| Android SDK 2.2/4.4 | Android software development toolkits |
| Star UML | Unified Modeling language tool |
| Microsoft Visio | Flow Chart software |
| Microsoft Words 2010 | Project documentation editor |

### 2.4.2 Hardware Requirement

Refer table 2.2 for hardware requirement in this project.

**Table 2.2: Hardware Requirement**

| Suggested Configuration | Minimum Configuration |
|---|---|
| 4GB RAM | 1GB RAM |
| Keyboard and Mouse | Any Keyboard and Mouse |
| Network Card | Any Network Card |
| Android Phone with KitKat | Android Phone with Froyo |

### 2.4.3 Other requirements

Refer table 2.3 for other requirement in this project.

**Table 2.3: Other Requirements**

| Suggested Configuration | Minimum Configuration |
|---|---|
| Android Supported Virtual machine | Any Android Supported Virtual machine |

## 2.5 Project Schedule and Milestones

Project schedule and milestones describes the action that need to be done throughout the whole project development. For every single project that in development process, conducting a project schedule and milestone are important. Both of them are useful in guiding the project flow so that the system/application can be delivered on time. In the project schedule, action is decomposed into every single task and the time frame is provided. Usually extra time is added in the project schedule. The developer must finished the task on time so that in won’t affect the whole process.

In this project, Grantt Chart is used to represent the project schedule and milestone. Tasks distribution with the timeline is schedule. Refer Appendix 1 for the Grantt Chart and Flow Chart.

## 2.6 Conclusion

In this chapter, literature review and methodology is done. Literature review is the summary of the research after exploration of a topic. It is crucial to study the current trend of existing product in the market to build an effective and higher quality of software product. On the hand, methodology consist the method to use throughout the whole project. A proper software development methodology will result to the successful in development process. This can be done by offering the best suit development framework for the development of the application. The listed project requirements do also give a basic overview on what are used to develop this project.

In next chapter, requirement analysis which describe both functional and non-functional requirement will be discussed. Those requirements will be show through diagrams and the system flow also will be explained later.

# CHAPTER III

# ANALYSIS

## 3.1 Introduction

In this chapter the requirements of the system is discussed. This includes the problem analysis which defines the requirements of the system, independent of how those requirements will be accomplished. This chapter includes requirement analysis which will be discussed later. In requirement analysis, module and function about this GeoTravel Application will be discussed.

Analysis represent the 'what phase'. In this phase, the problem that faced by the customer is define. Developer trying to solve their problems by developed a system. Deliverable result which is a requirement document is included. This document is needed at the end of this project to check whether the system meet the requirement that had set. In this document, it state clearly and precise the procedure and what to be built for the system. Requirement from the customer is captures and define the goals and interaction.

In this chapter, analysis phase start with problem analysis. Problem analysis phase normally describe the current system and follow by analysis the possible steps to improve the application. In this case, as far as I know, there is no similar system being developed so the real life scenarios of traveler become a way to analysis the traveler itineraries planning and turn it into business requirements of this application.

Besides that, functional requirement of this application in solving current problems will be discuss later. The non-functional requirements are explained too as the consideration of software quality issues. Other requirement such as software and hardware requirement which have been discussed in Chapter 2 also will further discuss in this chapter.

### 3.2 Problem analysis

As far as I aware, there is no similar automation travel system is being developed. Therefore, the real life scenarios will be discussed on finding the problems of current situation. In this project, how the travelers plan their itineraries will be the main consideration.

Traveling market is divided into two parts, casual customers and backpackers. For casual customers, normally they will pick a package from local travel agents. The package that provide by the travel agents usually is a pre-generated itinerary. The agency will provide all service in the trip such as book hotels, transportation and preorder the tickets. This prevents customers from constructing their personal itinerary and this save a lot of time. The aim for the agency is the first time travelers. Therefore the package provided by the agency normally just covers the most popular point of interests.

The problem faced by the travelers is the itineraries provided by the travel agents is lack customization and cannot satisfy individual requirements. Some of the point of interest in a place is missing due to agency preference. Besides that, package provided by the travel agents are too expensive for backpack travelers. Therefore, backpackers need to plan their trips in every detail include selecting the hotel, picking point of interests to visit, contacting the car rental service and so on.

In the real world, it is impossible for a travel agency to list all possible itineraries for customers to customize their itineraries. Therefore, a practical solution for this problem is to provide an automatic itinerary planning service. The user lists a set of interested point of interest and specific the time and budget, the service/application will returned itineraries to the user.

Algorithm that current use normally only consider a single's day trip. This is not a real case as in the real world most users will schedule multiday itineraries. Generating multiday itineraries is more complex than generating single day itineraries as in multiday itineraries, the point of interests cannot be repeated. To solve this, we can rank point of interests by their importance and use priority queue form the itineraries. Other than this, we also can exploit the geo-location by putting the point of interests which are nearby in the same day's itineraries.

Besides that, many travelers face the problem of provided point of interest is not the place that they want. This happen because the travel agents always provide the same set of trip plans, composed with point of interests but those point of interest may not attractive for the travelers who have visited the place before. Therefore, in this project, the selected point of interest is giving high priorities and generates the trip plan based on the priorities.

Therefore, in this system a novel itinerary planning approach is used. This method is to generate itineraries that narrow the gap between the agents and travelers. Basically in the preprocessing, the distance of two point of interest are evaluated by Google Map's API. List of point of interest provided in the map GUI to let the user to choose. The user can select preferred point of interest explicitly, while the rest points are assumed to be the optional. Different ranking functions are applied to different type of point of interest. The automatic itinerary planning service needs to return the itinerary with the highest ranking. Searching the optimal itinerary can be transformed into the team orienteering problem, which is an NP-complete problem without polynomial approximation. Therefore, a two stage scheme is applied.

In the preprocessing stage, parallel processing framework, MapReduce is used to iterate single-day itineraries. The results are maintained in the distributed file system and an inverted index is built for efficient itinerary retrieval. To construct a multiday itinerary, we need to selectively combine the single itineraries. The preprocessing stage, in fact, transforms the team orienteering problem into a set-packing problem, which has well known approximated algorithms. In the online stage, we design an approximated algorithm. In the online stage, we design an approximate algorithm to generate the optimal itineraries. The approximate algorithm adopts the initialization-adjustment model and a theoretic bound is given for the quality of the approximate result.

### 3.3 Requirement analysis

In this section, the behavior of the system is being capture. This include data requirement, functional and non-functional requirements.

#### 3.3.1 Data Requirement

In this project, database is used to store the data which needed for the system. The tools that used to store database is MySQL. The object of this system is recognized and the database is built based on the result. The objects of this system include information about point of interest, user location, and user history. The data should be gathered and well designed so that the system function well and integration between each other will be in a good quality and consistent.

In this project, the data to be displayed may come from more than one table which combined few tables in order to get the result. Some modules may not have its own table of data but it used a certain primary key from a certain table to pass on to the other modules. This make this application runs smooth when message passing.

#### 3.3.2 Functional Requirement

GeoTravel Application may be decomposed into several important components and each component interacts with each other to handle the data exchange and manipulation. Each component here provides its unique functionality to the system and may runs on different type of data from database.

These components can be called as sub-system and in this application they are divided into few units table 3.1.

**Table 3.1: Function Requirements**

| ID No. | Requirement | Description |
|---|---|---|
| GeoTravel_FR1_1 | Detect user location | The system should auto-detect the location of the user |
| GeoTravel_FR1_2 | | The system should display the point of interest based on the location of the user |
| GeoTravel_FR1_3 | | The system should notify user when there is no internet connection |
| GeoTravel_FR2_1 | Point of interest | The system should differentiate the point of interest. Examples of POI include historical place, event and restaurant. |
| GeoTravel_FR2_2 | | The detail of the POI should be show to the user when the user click the button |
| GeoTravel_FR2_3 | | The system should direct the user to the page that describe the place when the user clicking the location button |
| GeoTravel_FR2_4 | | The system should notify the user if the process is failed |
| GeoTravel_FR3_1 | Check the distance and the expected time | The system should show the distance between two point when the user click two point in the map |
| GeoTravel_FR3_2 | | The duration to travel along the point of interest should be display to the user |
| GeoTravel_FR3_3 | | The system should notify the user if the process failed |
| GeoTravel_FR3_4 | | The nearest route should be recommended to the users based on the point |

### 3.3.3 Non-functional Requirement

System constraints and performance expected in real life scenario are considered as a part of requirement to make sure the system to function as user expect and perform well. It is important to stress out the non-functional requirement to make sure the progress of functional business is being supported. The non-functional requirements are stated in table 3.2.

**Table 3.2: Non-functional Requirements**

| ID | Qualities | Description |
|---|---|---|
| GeoTravel_NFR1_1 | Availability | The system should be available to use most of the time. System operational time should be high. The system should also need to be available for service when users request it. |
| GeoTravel_NFR1_1 | Efficiency | The system should be efficient in providing the point of interests of the place and calculate the best route for the users. The system should fully utilize the resource such server cycle and memory. |
| GeoTravel_NFR1_1 | Flexibility | Flexibility specifies the expandability of the functionality of the system after it is deployed. The system should be flexible enough to extend the functionality after it is implemented. |
| GeoTravel_NFR1_1 | Reliability | The system should be reliable and all the components in the system should work properly. The capability of the software to maintain its performance over time is calculated. The failure rate of the system should be less than 1 over 100 requests. |
| GeoTravel_NFR1_1 | Security | Security ensures the integrity of the system from accidental or malicious damage and prevents unauthorized access to the system. The system should be secure from any unauthorized access as the information of the users is confidentiality. |
| GeoTravel_NFR1_1 | Platform Compatibility | This system should be able to run on any Android platform between API level 8 (Froyo) to API level 19 (KitKat) and with GPS location function in the device. |
| GeoTravel_NFR1_1 | Usability | The system should be easy to learned, understood and use by the user. The |
| | | system should also develop based on user requirement that is user-friendly. |

### 3.3.4 Others Requirement

- **Software Requirement**

**Table 3.3: Software Requirement**

| Item | Description |
|---|---|
| Microsoft Windows 7 | Operating system |
| Google Chrome/Mozilla Firefox | Web Browser |
| WordPress | Website creator |
| Eclipse Kepler | IDE for Android and Java development |
| Android SDK 2.2/4.4 | Android software development toolkits |
| Star UML | Unified Modeling language tool |
| Microsoft Visio | Flow Chart software |
| Microsoft Words | Project documentation editor |

- **Hardware Requirement**

**Table 3.4: Hardware Requirement**

| Suggested Configuration | Minimum Configuration |
|---|---|
| 4GB RAM | 1GB RAM |
| Keyboard and Mouse | Any Keyboard and Mouse |
| Network Card | Any Network Card |
| Android Phone with KitKat | Android Phone with Froyo |

### 3.3.5 Use Case

#### 3.3.5.1 Search Location

**Table 3.5: Search Location Use case**

| **Use Case Diagram** | Search Location |
|---|---|
| **Description** | User can search location based on point of interest or coordinate. The system will direct the map to the place the user search. |
| **Preconditions** | The system's point of interest table should have data stored and not to be empty |
| **Constraints** | Internet connection is require to direct the user to the location by showing in the map |
| **Normal Operations** | - From main menu screen, user can choose either enter place or latitude with longitude<br>- If the place that key in by user is not in the database, notification message will be display to the user<br>- If the place or point of interest that enter by the user is in the list, the system will show and direct to the location by using map<br>- Next, user can see the point of interest around that point |
| **Processing Frequencies** | None. User can search the location from everywhere and any time they want |

#### 3.3.5.2 Display Detail of POI

**Table 3.6: Display Detail Use Case**

| **Use Case Diagram** | Display Detail of the point of interest |
|---|---|
| **Description** | User can view the detail of the point of interest by clicking the location icon in the system |
| **Preconditions** | The system's point of interest table should have data stored and not to be empty |
| **Constraints** | Internet connection is require to direct the user to page that describe the detail of the point of interest |
| **Normal Operations** | - From the map that show the POI, user can click the POI that interesting to them<br>- The system will direct to the page and the details such as operation hour and admission fee is displayed<br>- If the system cannot search the detail of the POI, error message will be prompt |
| **Processing Frequencies** | None. User can search the location from everywhere and any time they want |

#### 3.3.5.3 Display the distance and duration between two point

**Table 3.7: Display distance Use Case**

| | |
|---|---|
| **Use Case Diagram** | Display the distance and duration between two point |
| **Description** | User can view the distance and duration of two point |
| **Preconditions** | The system's should get the point of interest from the user |
| **Constraints** | Internet connection is require to direct the user to page that describe the detail of the point of interest |
| **Normal Operations** | - From the map that show the POI, user can click two POI that interesting to them<br>- The system will calculate and show the shortest distance and the duration between two point<br>- If the system cannot search the point of interest that user enter, error message will be prompt out |
| **Processing Frequencies** | None. User can search the location from everywhere and any time they want |

## 3.4 Conclusion

Analysis phase is important for gathering the system requirement so that the functionalities of the system fit to the scope of the system. Through this chapter, functional requirement, non-functional requirement and use case of the system is covered. These requirements are important in order to develop a system which fulfilled user requirement. At the end of the development phase, user will be test the system based on the requirement, therefore the system should fulfill user satisfaction. This chapter also constructed to improved current scenarios to be better.

In next chapter, the design and the technique used to develop this project will be discussed. This will include the High Level Design (HLD), Detail Design (DD) and the technique used in the analysis part.

# CHAPTER IV

# DESIGN/THE PROPOSED TECHNIQUE

## 4.1 Introduction

In this chapter, the system design and function will be discussed. Besides that, this chapter will also include the proposed technique which will be used in the analysis phase. System design in this chapter will be discussed based on high-level design and high level design based on the analysis that had discussed in the previous chapter. System architecture, user interface and database design is included in the high level design.

This chapter is important to give a clear picture about how the system functions. In order to determine reliability of this system, the design phase should be included high level design. This chapter can be also used as a reference manual that stated how the system and module interact in the real environment.

## 4.2 High-Level Design

This section discussed the high level design of the project which includes architecture design, class diagram, screen design and database design. Function of high level design is to analyze all requirements that are needed for the software. The structures of the high level design which must be fulfilled by the developer are functional and non-functional design.

### 4.2.1 Software Architecture

The software architecture is the set of structures about the system. It included software elements, external visible properties and the relationship between the components or module. The main purpose of this section is to asses and evaluates the system. The software architecture is frequently used as a means to manage the configuration of the product.

### 4.2.2 Quality Requirements

GeoTravel application is planned to be recommend the traveler point of interest in the place that traveler wish to visit. This will be a public application and the attributes that used in this system are performance, availability, reliability, consistency, modifiable and traceable.

**Table 4.1: GeoTravel Application's Quality Requirements**

| | **Attributes** | **Details** |
|---|---|---|
| 1 | Performance | Response to queries should take no longer than 5 seconds to load the result or data onto the screen to be display to user after user submitted the query. |
| 2 | Availability | The system should available all the time. |
| 3 | Reliability | The failure rate of the system should be less than 1 over 100 requests. |
| 4 | Consistency | The system requirements must be consistent and do not conflict with each other. |
| 5 | Modifiable | The system should be able to modified or make changes. |
| 6 | Traceable | All the data should be able to trace back. |

### 4.2.3 System Architecture

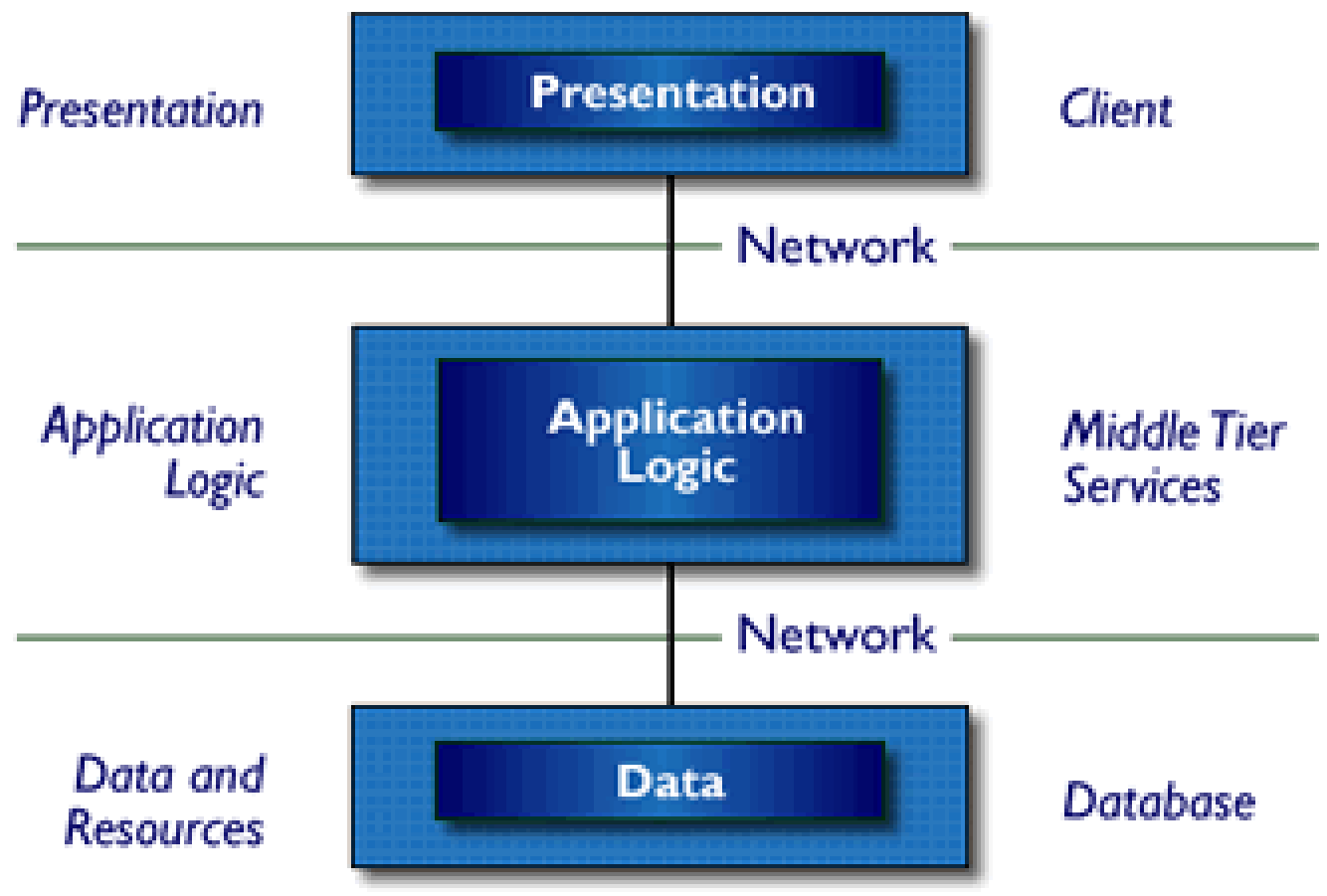

**Figure 4.1: 3-tier Architecture**

The system architecture for GPS Trajectories system is illustrated in the above figure. It represented in three-tier architecture which consist of Client tier (presentation), Middle Tier (application and logic) and Database (data and resources).

The architecture of GeoTravel application included server, database and device. User use a device to access and view the system that supported by a server. When the user enter a query, the server send the request to database and the database process the query and send the result to be display to the user on the device.

Flow chart of the system is shown in the figure 4.2.

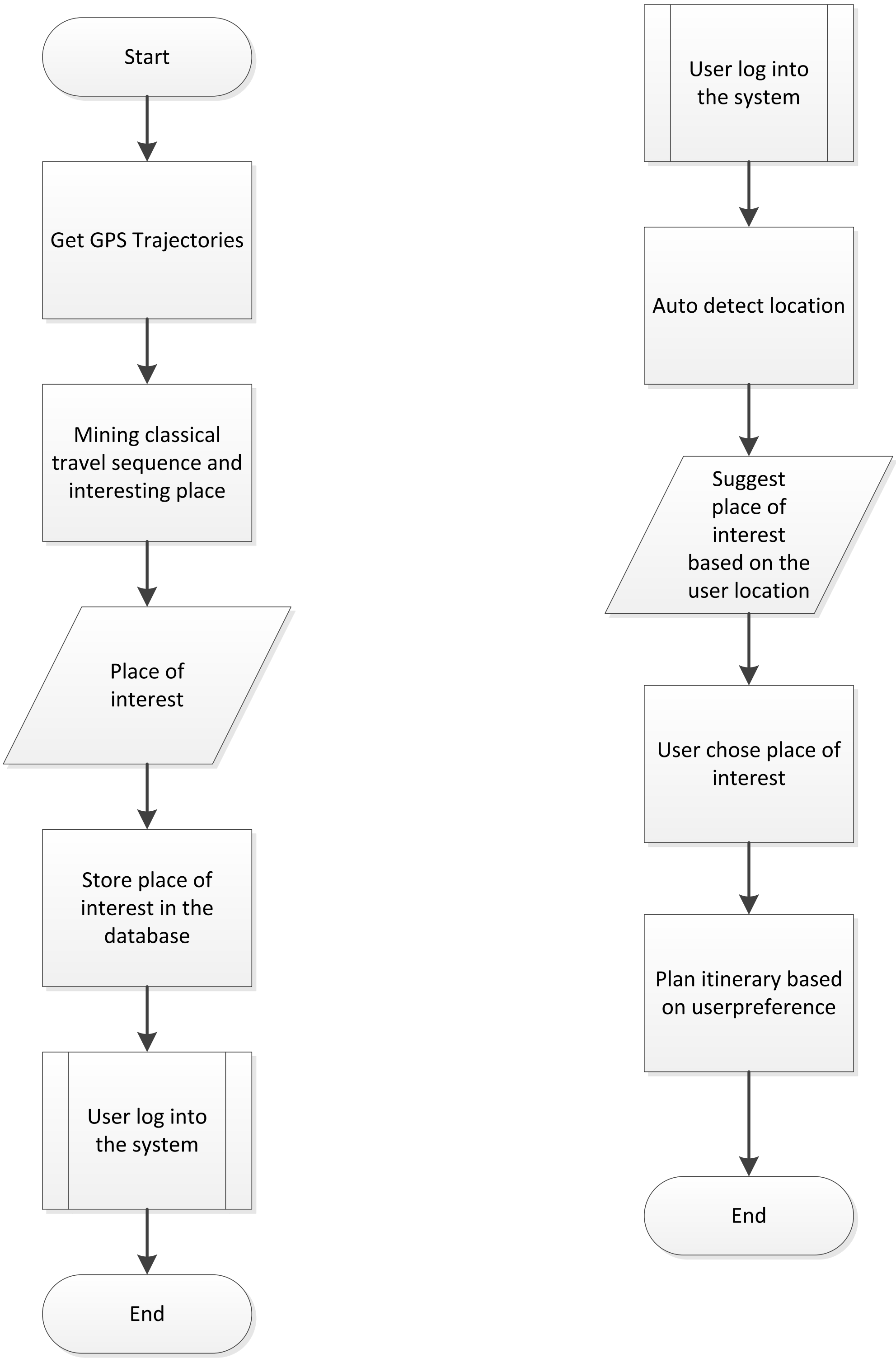

**Figure 4.2: Flow Chart**

### 4.2.2 User Interface Design

User interface design is the screen design of the system on how a user will be able to interact or use the system. User interface design is important to show a more attractive system. Besides that, the user interface may influence the user to understand about the system. A good user interface will help the user more understanding about the system and thus easier to use it. Cost for modify the system at the end of the project will be reduced if the screen design is plan well at the beginning.

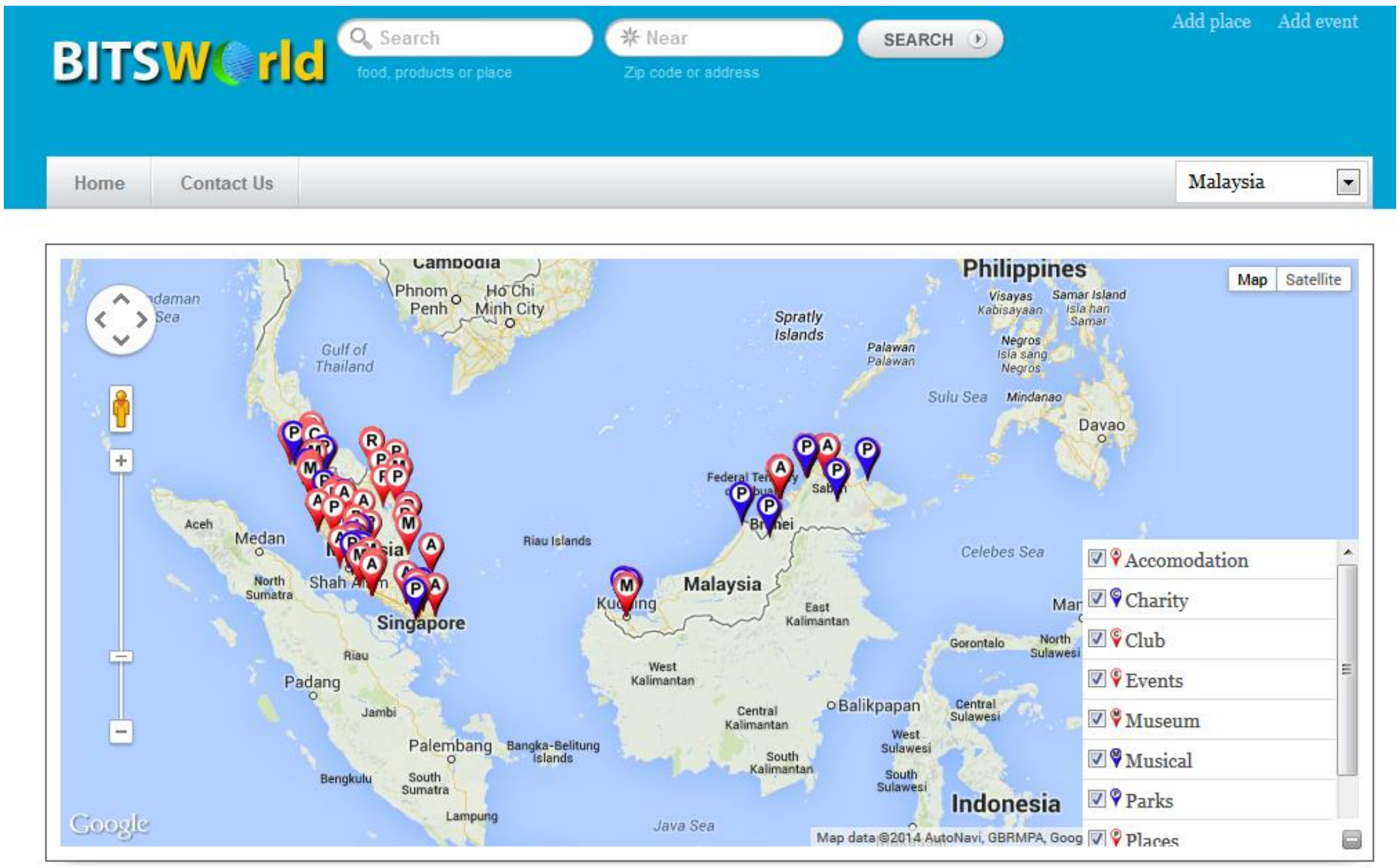

**Figure 4.3: Screen Design for the web**

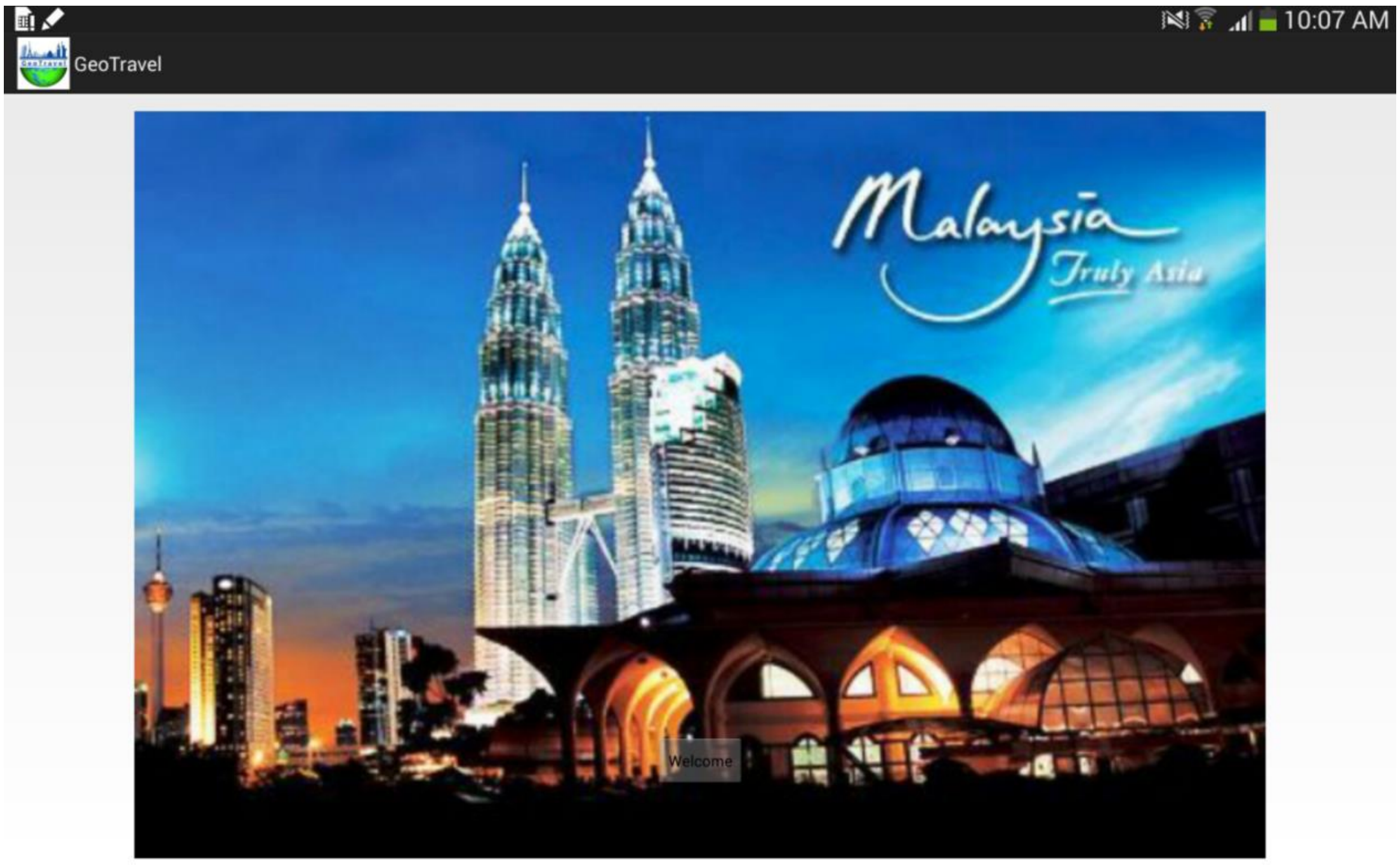

**Figure 4.4: Screen Design for Main Menu**

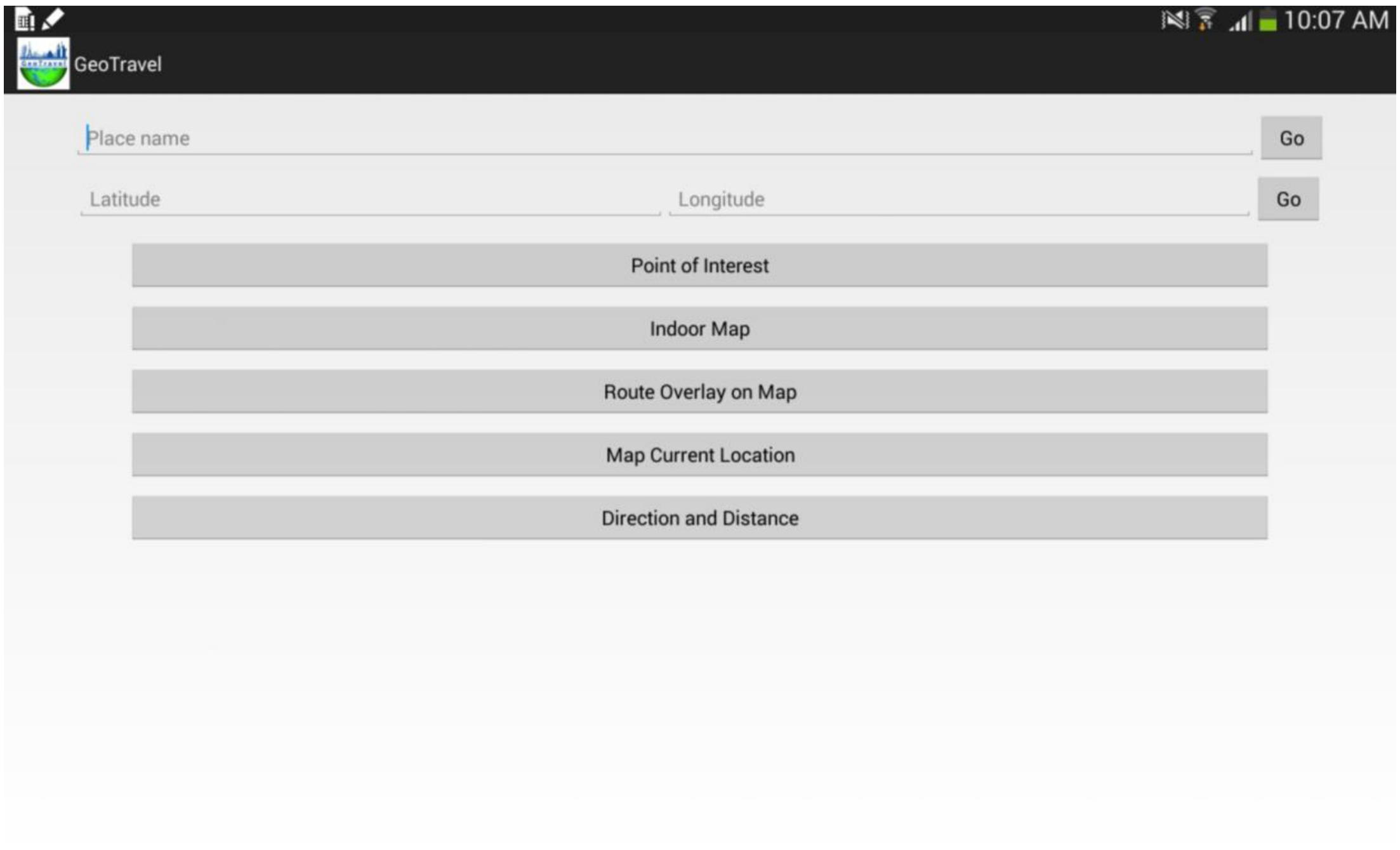

**Figure 4.5: Screen Design for Menu List**

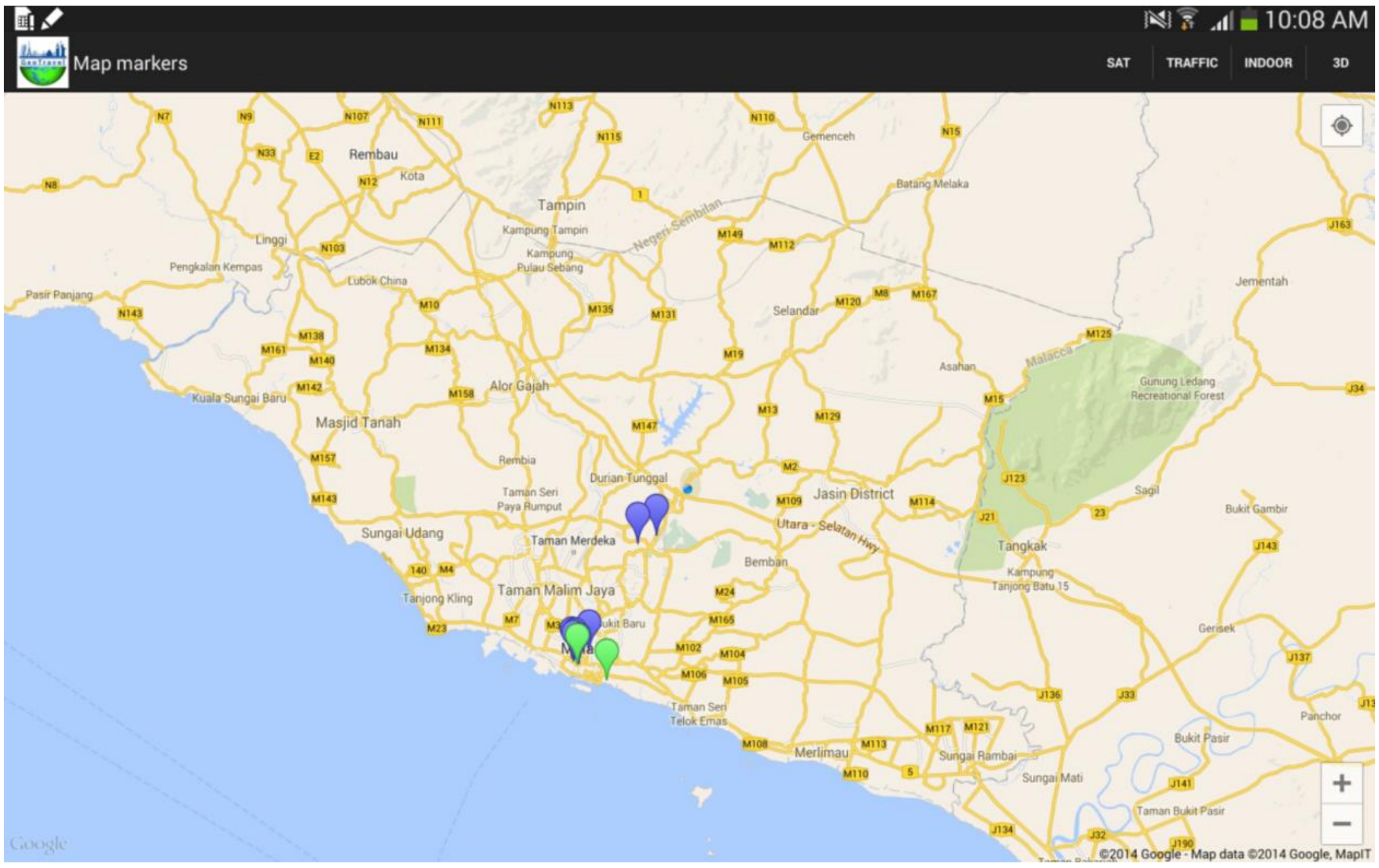

Figure 4.6: Screen Design for Point of Interest

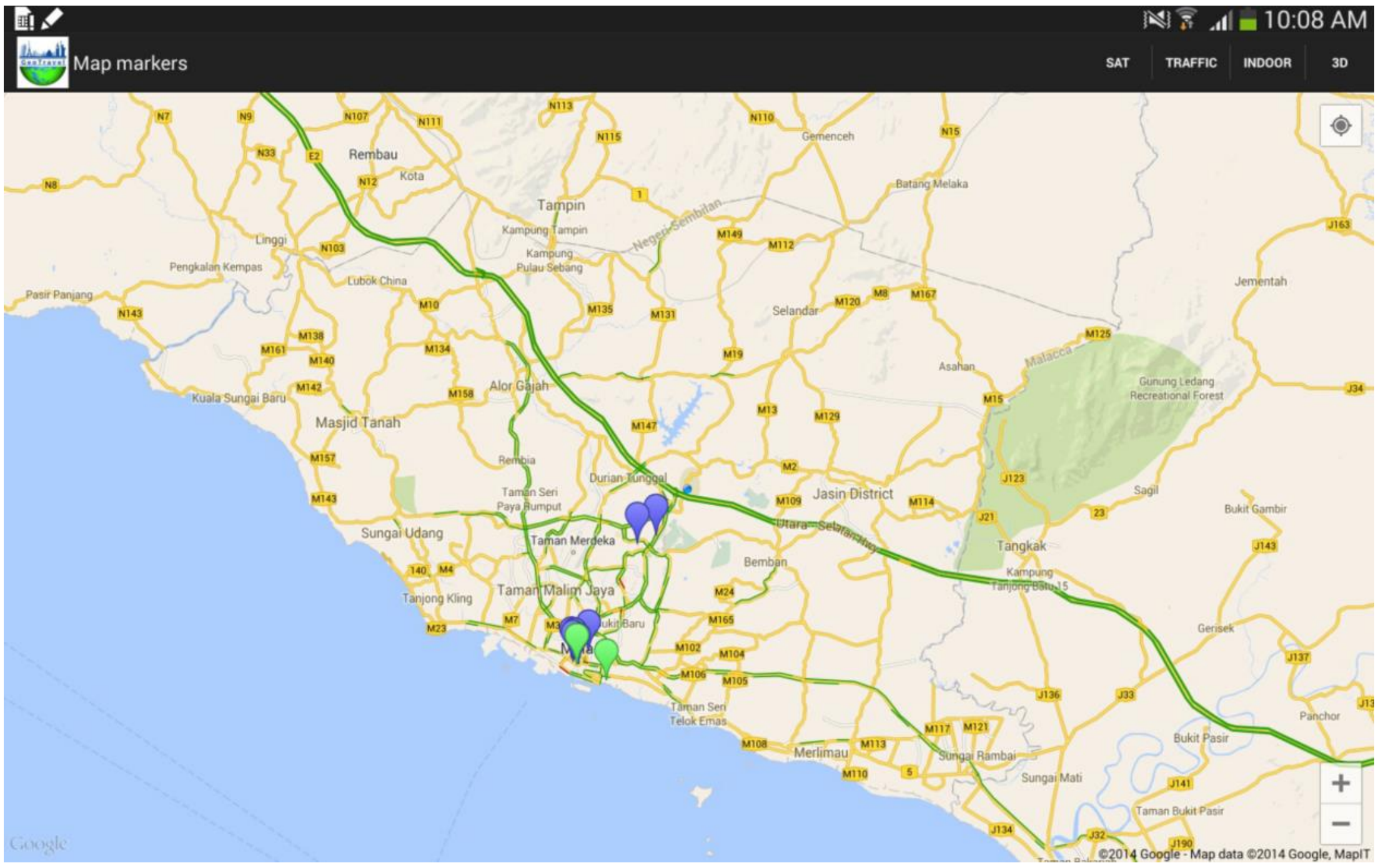

Figure 4.7: Screen Design for Traffic Analysis

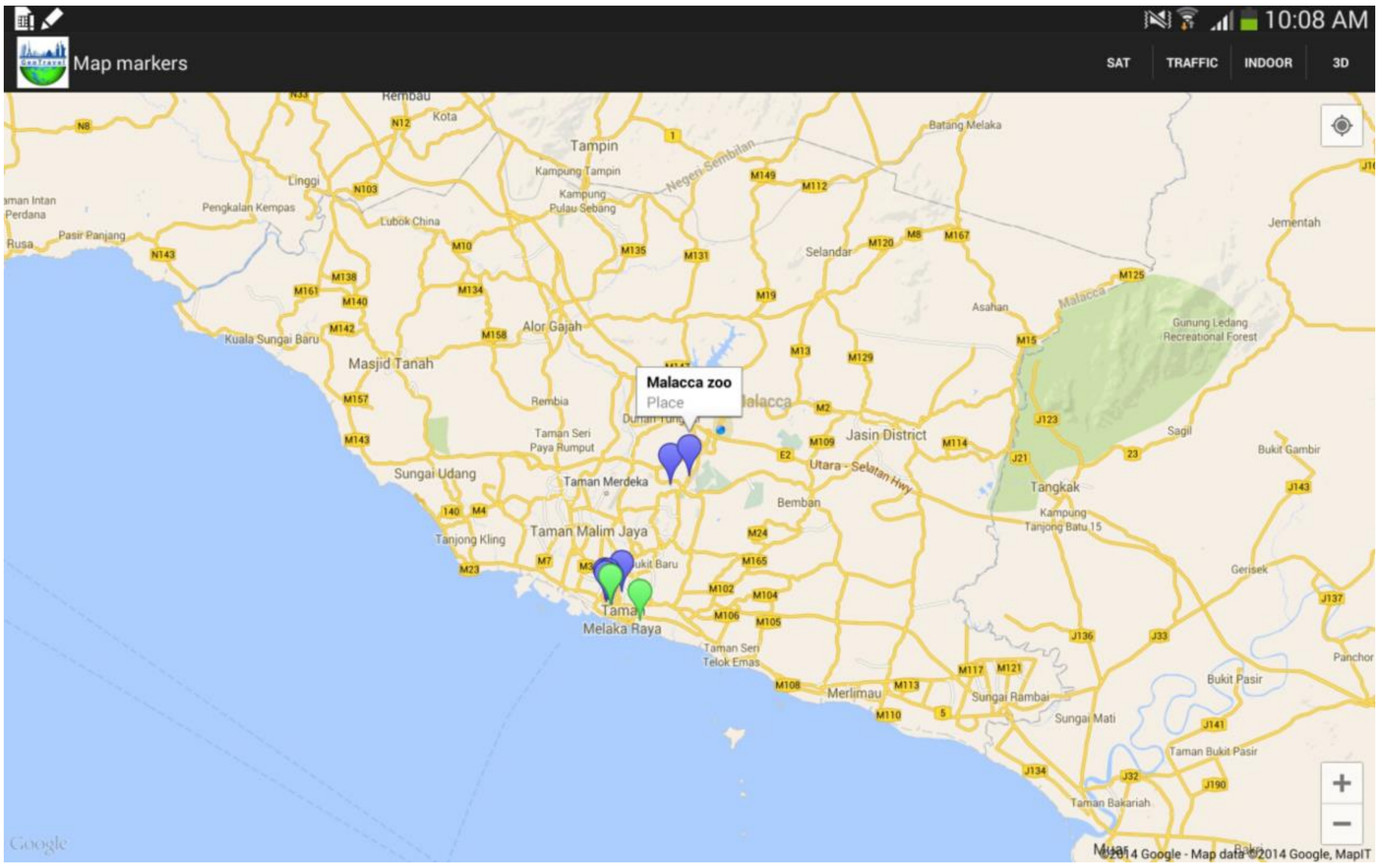

**Figure 4.8: Screen Design for Detail**

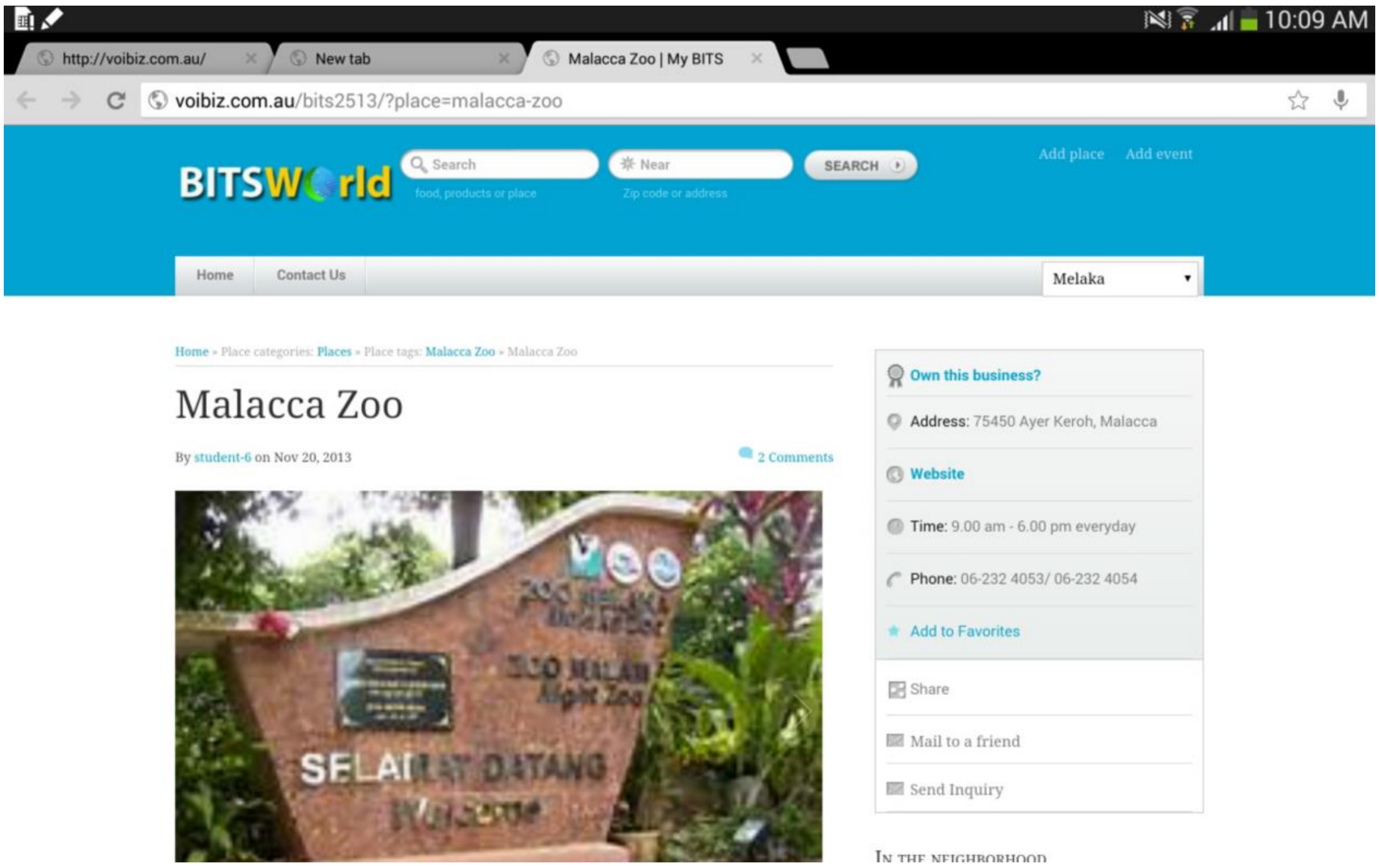

**Figure 4.9: Screen  Design for the description of POI**

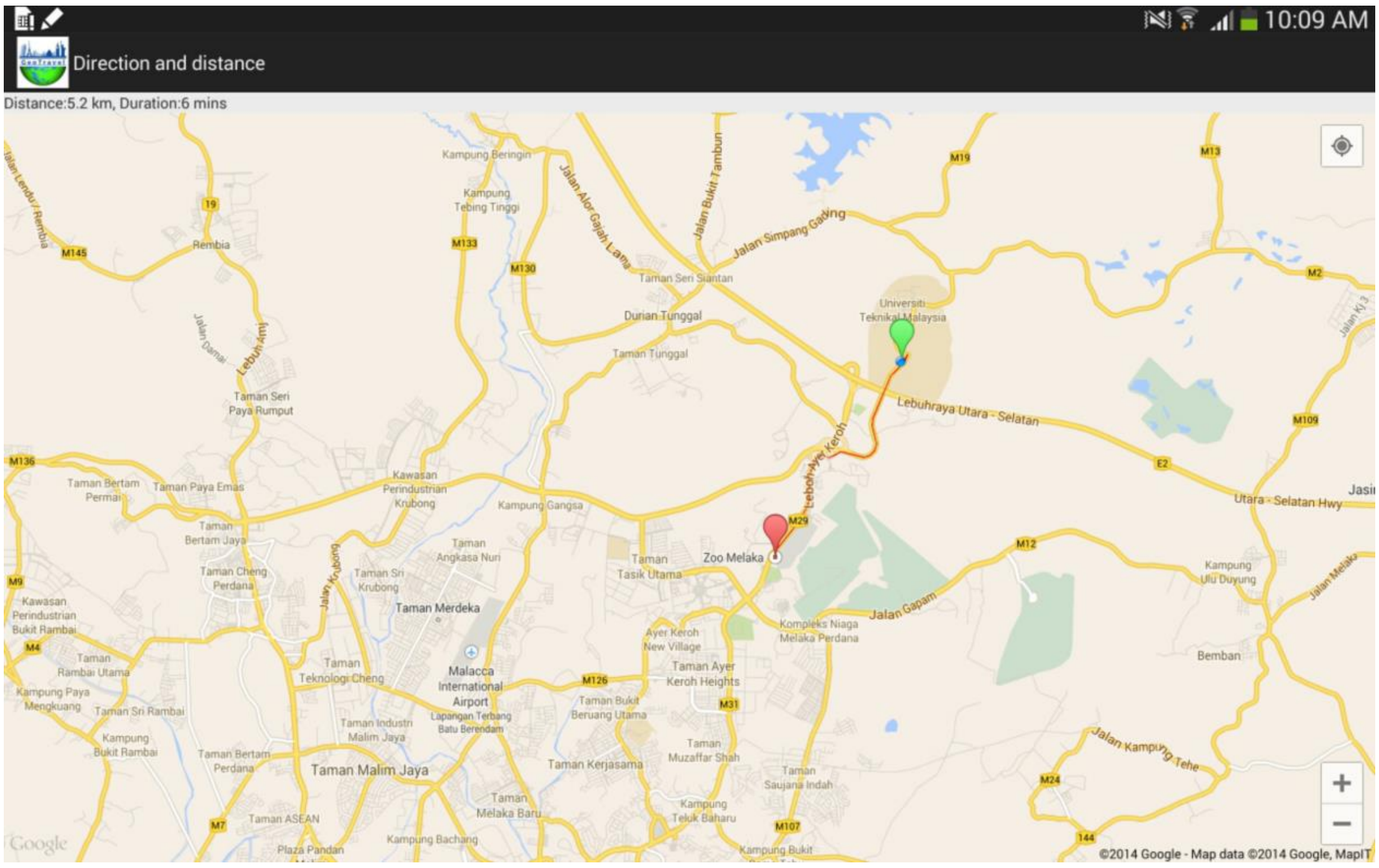

**Figure 4.10: Screen Design for Distance and Duration**

#### 4.2.2.1 Input Design

Input design emphasize on the entry of data into the system. The system need to receive input from users in order to process its functions. The interfaces and forms are to be designed to get input from users.

#### 4.2.2.2 Output Design

Output design is the design that emphasizes on presenting the information retrieved from the system database based on user's request.

### 4.2.3 Database Design

Database design is the design of interaction of data entity in the system. The information that needed to run the application is store in the database. Therefore a well design database will improve the efficiency of the application. Database design is included in the next section.

### 4.2.3.1 Conceptual and Logical Database Design

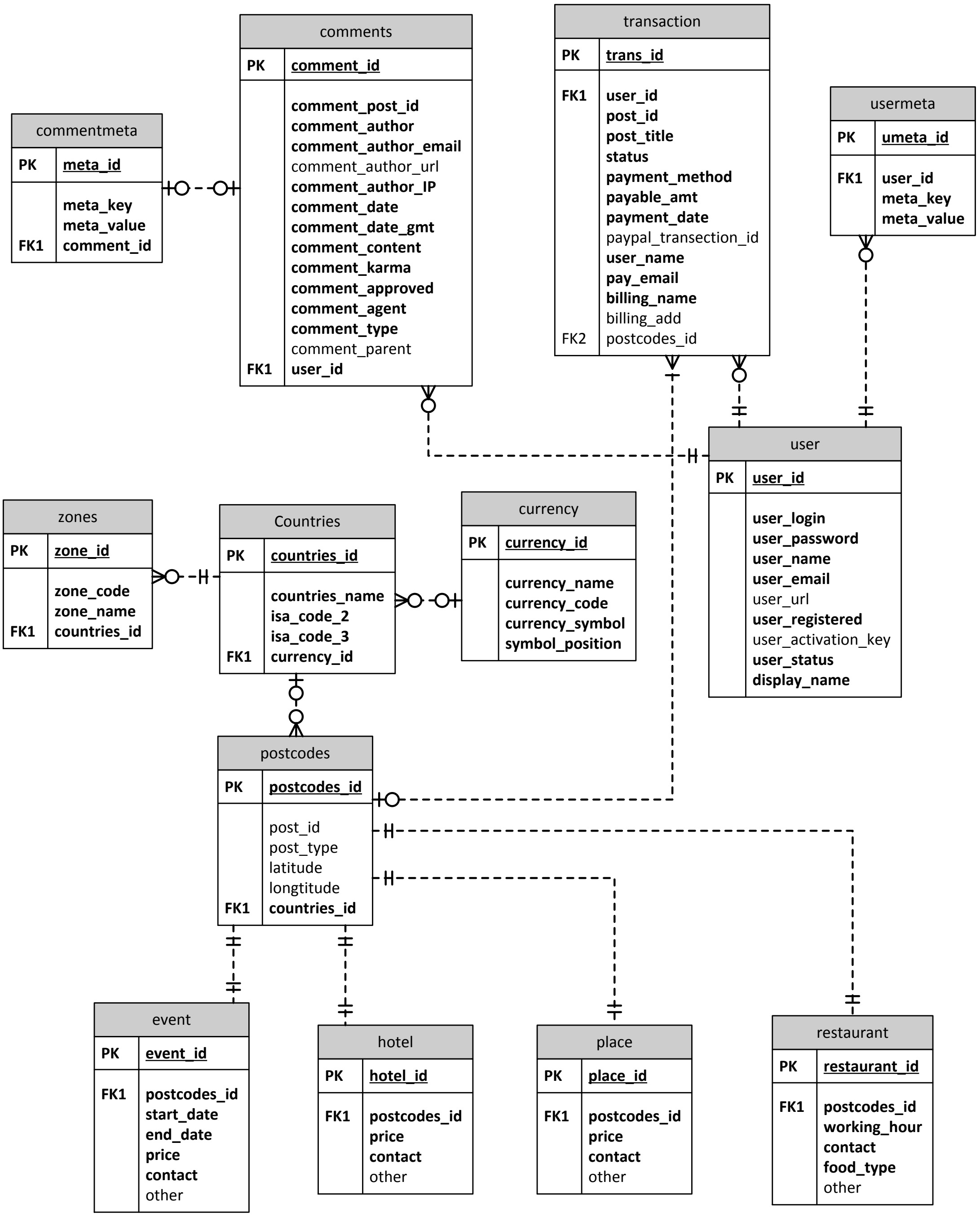

**Figure 4.11: ERD Design**

### 4.3 Detailed Design

After high level design, documentation focus is on detailed design which is low level design. In detailed design, each module's responsibility should be specified as precisely as possible.

#### 4.3.1 System Specification

**Table 4.2: System Specification**

| **Title** | **Description** |
| --- | --- |
| Project | GeoTravel Application |
| Sub-system name | none |
| Author | Liew Li Ching |
| Program name | GeoTravel |
| Description | This application is to recommend point of interest to traveler |

Classes used in this application are:

1. MainActivity.java
2. MainMenu.java
3. RouteMapper.java
4. MapMarker.java
5. Direction.java
6. Duration.java
7. Setting.java

### 4.4 Technique Propose

For the analysis part, data is collected through Garmin GPS website. GPS trajectories about Melaka are collected and stored. The architecture for mining the interesting location and traveler travel sequence are as below:

i. GPS trajectories/logs is collected
ii. Pre-processing is done before mining process
iii. Detect the stay point
iv. Mining the location interest by using OPTICS Density Based Clustering
v. The point of interest is show as the result of the mining process

#### 4.4.1 Data Collection

The main data used in the analysis part is GPS trajectories of various users on different date and time. The GPS trajectories data is collected from Garmin GPS which are in text file. After that, the data collected is converted to .gpx file so that it can map the trajectories on the map.

#### 4.4.2 Pre-processing

In pre-processing, the data collected is pre-process before the mining process. The method used is cleaning process. During the cleaning process, the data that is improper is being removed. The data that have the speed value more than 300km/h is removed. This is because that is impossible for a car or people to travel at 300km/h except they are traveled by plane. This may be due to malfunction of the GPS.

#### 4.4.3 Stay Point Detection

After pre-processing process, the stay point is detected by using below algorithm. A simple system is done to detect the stay point base on the algorithm. The stay point is detected if the distance is not more than 200 m and the time exceed threshold time which is 20 min. Then the point is a stay point. Haversine' formula is used to calculate the distance between 2 points.

$$d = 2r \sin^{-1}\left(\sqrt{\sin^2\left(\frac{\emptyset i - \emptyset j}{2}\right) + \cos(\emptyset i)\cos(\emptyset j)\sin^2\left(\frac{\varphi i - \varphi j}{2}\right)}\right)$$

**Figure 4. 12: Harversine' Formula**

Input : GPS trajectories
Output: Stay Point

BEGIN
    For each GPS trajectories, do
        Traj=LogParsing(P)
        S=StayPointDetection(Traj, $Distance_{threshold}$, $Time_{threshold}$)
        StayPoint.Add(S)
    End
END

Above is the algorithm to determine the stay point. The stay point is detected and removed the inconsistent point. After this, a Tree Based Hierarchical Graph (TBHG) is plotted based on the stay point.

After getting get the TBHG, the classical travel sequence of the users is mined.

### 4.4.4 Mining of Stay Point

In mining process, clustering method is being used. Clustering of stay point is to get the point of interest which is the common places for traveler. The clustering method used is OPTICS density based clustering. This clustering is based on density of the data. This mean that the stay point is clustered based on the location. The more popular place will have higher density.

## 4.4 Conclusion

Design of the architecture is the important part in a system. Useful of system is depends on the architecture. Therefore, in this chapter design of architecture is discussed clearly with the database, system design, detailed design and etc.

In the next chapter, implementation of the design system will be discussed.

# CHAPTER V

# IMPLEMENTATION

## 5.1 Introduction

In this chapter, implementation phase of the project will be discussed. This include software and hardware development environment, configuration management and version control procedure that being used in this project. The purpose of this phase is to develop and deliver product to client after testing phase. Installation, setup, coding and testing is the process in this phase.

Software and hardware development environment will be discussed in the first section. This project involved environment requirements like Java Kepler environment to develop android application and server environment. The environment setting must be setup before the developer can start implementation phase.

The next section is this chapter is about software configuration management. In this section configuration management plan is discussed. This included the version on the documentation, user access right, system backup management and etc.

At last, the implementation status for each module is summarised. The details show in the module included the module size, the start date and the end date of the module.

## 5.2 Software or Hardware Development Environment setup

Java Development Kit (JDK) must be done at the beginning of the phase because JDK is important in developing android application using Eclipse. Besides that, server setup and runtime environment setup are also important for this project. Therefore, this setup must be done before the development activity start.

### 5.2.1 Java Environment Setup

Java Development Kit (JDK) and Java Runtime Environment (JRE) are important to run Eclipse IDE. Without JDK and JRE, the development will failed to compiled and run.

Download the package of Java EE 6 which included JDK 7. To setup the JDK in a computer, the following steps are needed:

- Download the Java EE 6 zip file
- Go to ***setup/installers***
- Unzip ***java_ee_sdk-6u4-jdk7-windows-x64-ml.zip*** in your computer
- Run ***java_ee_sdk-6u4-jdk7-windows-x64-ml.exe*** file. The file will automatic run in cmd.exe when click it
- The installation wizard will pop up
- Follow the instruction on the installation wizard
- Refer http://www.oracle.com/technetwork/java/javaeejavaee6sdk -install-jsp-140358.html for more details of installation process

### 5.2.2 Web Application Server

Developing a web application need a server to host the web application. There are a lot of servers hosting available in the internet. Apache Tomcat 7.0 will be used as the server in this project.

Step to setup Apache Tomcat 7.0 is as below:

- Download Apacthe Tomcat 7.0 on Apache Web
- Unzip ***apache-tomcat-7.0.29.zip***
- Run ***apache-tomcat-7.0.30.exe***
- The installation wizard will pop up
- Follow the instruction on the installation wizard.
- Refer http://www.mulesoft.com/tomcat-windows for more informatio

### 5.2.3 Software Development Tools Setup

Installation process of software development tools that required for developing GeoTravel application.

#### 5.2.3.1 Eclipse IDE

Eclipse IDE for this project is using Java EE Developers (Kepler Packages) . The purpose of IDE is to develop an application which include coding and testing.

Follow step below to setup Eclipse IDE (Keepler Packages):

- Download the Keepler Package from internet
- Go t***setup/installers/***
- Unzip ***eclipse-jee-kepler-SR2-win32-x86_64.zip***
- Click ***eclipse.exe*** and the eclipse is run on your computer
- Refer http://www.eclipse.org/resources/?category=Getting%20Started for more details

#### 5.2.3.2 Android Development Tools (ADT)

Android Development Tool (ADT) is a plugin that needed to build Android application. This plugin can be installed through Eclipse.

Follow the step below to install the plugin:

- Run Eclipse on your computer
- Select Help, then choose Install New Software.
- On the right corner, click add button
- Enter https://dl-ssl.google.com/android/eclipse for "ADT Plugin"
- Click OK.
- Follow the wizard.
- Choose the tools you wish to download, the click next
- Click accept liscense agreenment if you agree with the liscense
- Click next and it will start download.

## 5.3 Software Configuration Management

### 5.3.1 Configuration Environment Setup

**Table 5.1: Environment Setup**

| LC Stage | Deliverables | Responsible |
|---|---|---|
| Planning | • Planning Document | Onsite |
| Requirements Analysis | • Software Requirement Specification Document | Onsite |
| Design | • High Level Design<br>• Detailed Design Document | Onsite |
| Build | • Source Code<br>• Unit Test Result | Onsite |
| Integration Test | • Integrated Software<br>• Integration Test Result | Onsite |
| Documentation | • User Manual | Onsite |

### 5.3.2 Version Control Procedure

**Table 5.2: Version Control**

| Section | Version Type | Revision Number |
|---|---|---|
| 1 | Draft | 0.0a |
| 2 | Review | 0.0b |
| 3 | Minor Changes | 1.1 |
| 4 | Major Changes | 2.0 |
| 5 | Baseline | 1.0 |

#### 5.3.2.1 Access Right

**Table 5.3: Configuration Access Right**

| Type of area | Access rights |
|---|---|
| Draft | Read, Write, Delete, Destroy |
| | Read, Write, Delete |
| | Read, Write |
| Review/test | Read, Delete, Destroy |
| | Read |
| Baseline | Read, Delete, Destroy |
| | Read |
| Release | NA |

#### 5.3.2.2 Backup Management

**Table 5.4: Backup Management**

| Storage Area | Backup Media | Backup | Checking backup files |
|---|---|---|---|
| | | Frequency | Frequency |
| All | Hard Disk | Weekly | Monthly |

## 5.4 Implementation Status

Implementation status stated the progress for each module in term of module planning. The time or duration of each module is stated and the date complete of each module is recorded.

**Table 5.5: Analysis Status**

| Step | Duration to Complete | Date Completed |
|---|---|---|
| Data Collection | 10 | 12 Jun 2014 |
| Proposed Technique | 10 | 22 Jun 2014 |
| Analysis | 5 | 27 Jun 2014 |
| Other | 3 | 30 Jun 2014 |

**Table 5.6: Implementation Status**

| Module | Duration to Complete | Date Completed |
|---|---|---|
| Main Menu | 3 | 3 July 2014 |
| Point of interest | 4 | 7 July 2014 |
| Traffic Analysis | 5 | 12 July 2014 |
| Route Suggestion | 3 | 15 July 2014 |
| Analysis Result | 2 | 17 July 2014 |
| Others | 2 | 19 July 2014 |

## 5.5 Conclusion

This chapter is about implementation phase in project development life cycle. It explained about process to development system which included setup before code. Tasks like software or hardware development environment setup, configuration environment setup, and implementation status is discussed in this chapter.

In the first section of this chapter, software or hardware development environment setup is discussed where environment setup process is discussed in detail. These setup are important and must be done before develop process so that coding can be implemented on time. Environment setup included setup of software included java environment setup, database server setup, web application server setup, Eclipse setup and android development setup which are needed in development and testing process. In this project, GeoTravel application can be run either on web or android platform. Therefore the mobile server and web server must be setup before GeoTravel application can be run and tested fully.

In the next section, detail of software configuration management is discussed. Configuration of environment setup and version control procedure are recognized and documented. The flow and the version is determined and recorded so that it gives clear information of the version. In configuration plan, the flow between working stage, review stage, and finally stage have their own version of numbering and must be follow throughout the project.

Lastly, implementation status of this project application is discussed in the last section. In this section, analysis status and the implementation status for module is conducted. The duration of complete and date completed must be followed so that the project schedule won’t be affected and can deliver on time. The duration to complete the analysis and implementation status is roughly 5 weeks.

# CHAPTER VI

# TESTING / EXPERIMENTAL RESULTS AND ANALYSIS

## 6.1 Introduction

This chapter is about testing, experimental results and analysis. This includes test plan, test environment, test strategy, and analysis experimental result. In testing phase, the system is tested and checked function by function so that the developer can detect the fault of the system. Based on the testing result, developer can fixed the error before establish the system to public.

In test strategy, approach and architecture to be used to perform testing is discussed. The purpose of a test strategy is to clarify the major tasks and challenges of the test project. Test Strategy included black box testing, cause-effect, graphing, boundary testing, and white box testing to test this product against its specification. In this project, only bottom-up, white box testing and black box testing will be conducted.

Test plan is the document that describing the testing scope, activity and perform a test to the system. Any software product or application project are required to undergo test plans. Test plans are prepared for every stage of the system so that tester can test the whole system. Test plans included unit testing, integration testing, and user acceptance testing. Every stage of test plan has its own function and format to be followed.

Besides that, this chapter also included test design which is the design or planning on test phase that is suitable for testing this application. In this section, test plan is explained in detail which including, test name, condition and expected result. The purpose of test design is to make sure the tester can test the application systematically and perfectly by follow the plan.

After perform test case, the test result will be documented as test result. Test result included the error or failure of the application. If error is get as the test result, the detail of error or failure must be stated clearly in the document.

Last section of this chapter is analysis and result. This included result of mining GPS trajectories.

## 6.2 Test Plan

Test plan is the document that describing the testing scope and activity of application thus perform a test to the system

### 6.2.1 Test Organization

#### 6.2.1.1 Integration Test Plan

##### 6.2.1.1.1 Integration Test Environment

- **Hardware**

  Laptop

  i. ASUS N43, Intel (R) Core(TM) i5-2410M CPU @ 2.30GHz, 4.00GB of RAM
  ii. Mouse
  iii. Keyword

- **Software**

i. Window 7 Ultimate Service Pack 1
ii. Eclipse Keepler
iii. Android SDK
iv. Google Chrome

- **Communication**
  - Internally

- **Security Level**
  - At each module

- **Specific test needs**
  - None

#### 6.2.1.1.2 Integration Parameter

- **Critical Modules**
  i. Main Menu
  ii. Search Location
  iii. Point of Interest
  iv. Traffic Analysis
  v. Distance and Duration

- **Interfaces among module**

- Main Menu → Search Location/ Point of Interest/ Traffic Analysis/ Distance and Duration/ User manual

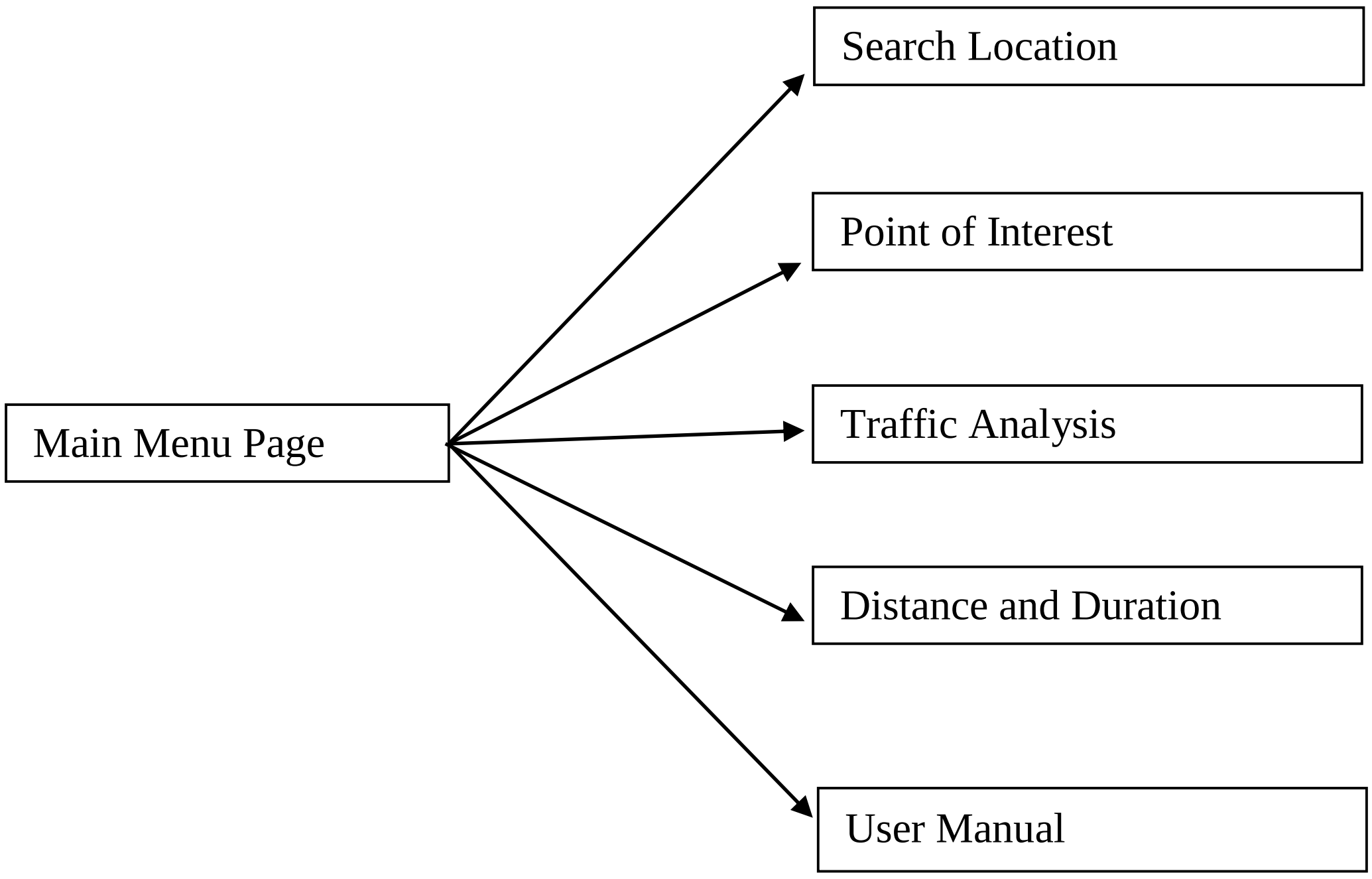

**Figure 6.1: Interface among Module**

- **External interfaces**
    - None

**6.2.1.1.3 Integration Procedure**

- **Order of integration**

Main Menu →Search Location →Point of Interest →Traffic Analysis →Distance and Duration

- **Activities, techniques, tools**
    - None

- **Test Execution procedures**

- Input data manually into form fields.

- **Test result checking method**
  - Visual inspection and comparison to the expected result

- **Test stop criteria.**
  - Mandatory error

- **Integration Test report**

**Table 6.1: Integration Test Report**

| | |
|---|---|
| **Author** | Liew Li Ching |
| **Tested By** | Liew Li Ching |
| **Subsystem Name** | **-** |
| **Funtionality Test** | GUI |
| **Test Data** | refer IntegrateTestPlan.xls |
| **Test Condition** | IntegrateTestPlan.xls |

**6.2.1.2 Unit Test Plan**

**Table 6.2: Unit Test Plan**

| | |
|---|---|
| **Author** | Liew Li Ching |
| **Application** | GeoTravel Application |
| **Tested By** | Liew Li Ching |
| **Subsystem Name** | **-** |
| **Funtionality Test** | All module |
| **Test Execution** | 1. Main menu<br>2. Search Location<br>3. Point of interest<br>4. Traffic<br>5. Distance and Duration |
| **Test Data** | refer IntegrateTestPlan.xls |
| **Test Condition** | IntegrateTestPlan.xls |

### 6.2.2 Test Environment

- **Hardware**

  i. Samsung Galaxy NOTE 10.1
     - Android 4.3 Jelly Bean
     - Memory: 3GB RAM, 16GB
     - 1.9GHz Quad + 1.3GHz Quad

  ii. Sony Ericson Xperia X10
     - Android 2.3.3 GingerBread
     - Memory: 384 MB RAM, 8GB
     - 1 GHz Scorpion

- **Software**

  i. Microsoft Windows 7 Ultimate Service Pack 1
  ii. Eclipse Kepler
  iii. SDK environment
  iv. Google Chrome

- **Test Stop Criteria**
  - Mandatory error

### 6.2.3 Test Schedule

**Table 6.3: Test Schedule**

| **Date** | **Module** | **Result** |
|---|---|---|
| 1 August 2014 | Main Menu | Functioning |
| 4 August 2014 | Search Location | Functioning |
| 8 August 2014 | Point of Interest | Functioning |
| 12 August 2014 | Traffic Analysis | Functioning |
| 15 August 2014 | Distance and Duration | Functioning |

## 6.3 Test Strategy

Test strategy in a project is important to reduce the testing time. Test strategy is used to produce an efficient test case so that the application can be test efficiently in order to improve the quality of application.. The purpose of a test strategy is to clarify the major tasks and challenges of the test project. Test Strategy included black box testing, cause-effect, graphing, boundary testing, and white box testing to test this product against its specification. In this project, only bottom-up, white box testing and black box testing will be conducted.

Bottom-up testing approach is used in this project. The lowest level components are tested first follow by higher level components. It starts from testing the small component in each modules and follow with integration test which is tested in integrated module. The bottom-up testing is end up with testing the whole system which user acceptance test is performed. The reason of choosing bottom-up approach as test strategy in this project is because this application has 5 modules and each module has to complete their own functionality so that the system can be work together.

White box testing also known as structural testing is the testing method which used to test the internal mechanism of the system. It is a verification techniques used by the developer to examine whether their code works as expected. Usually, white box testing involved unit testing and integration testing.

Black box testing which also known as functional testing is the testing phase that ignores the internal mechanism of system or component. Black box testing only focused on the output that generated in response. User acceptance testing is one of the examples of black box testing as users only test the external function of the system and do not know what is the code structure and how the internal work.

## 6.4 Test Implementation

Test implementation is the process where test condition is transformed into test case. At this phase, test environment is setup to perform test on the application.

Major task in test implementation are:

i. Develop test case
ii. Collect test data
iii. Create test data
iv. Write test procedures
v. Create test from test case
vi. Verify test environment
vii. Execute test case
viii. Compare result

## 6.5 Data Analysis

### 6.5.1 Experimental / Test Description

The data that used in this project is real-time data. Collection of data is done by collect the real time trajectories from the users. In the real case, the data needed is huge which may cause one year to collect. Therefore, to make it simple, this project only focused on Melaka state. There are 10 users involved and each user contributed more than ten trajectories which they travel along Melaka. Each trajectory contained around 100 GPS coordinate.

After collect all the data, process data mine the interesting location and classical travel sequences in a given geospatial region is done. In this process, tree-based hierarchical graph (TBHG) is used to model multiple's individual's location histories. Hypertext Induced Topic search (HITS) based inference model is used after TBHG process. This model infers interest of a location depends on not only the number of users visiting this location but also users' travel experience.

After mining the point of interest, novel itinerary planning approach is suggested in this project to help the user plan their itinerary based on their preferences.

### 6.4.2 Data

Sample Data is show in table 6.4.

**Table 6.4: Sample Data**

| Latitude | Longitude | Time |
|---|---|---|
| 2.19711 | 102.2487 | 4:09:22 |
| 2.19705 | 102.2487 | 4:09:27 |
| 2.19698 | 102.2487 | 4:09:32 |
| 2.19696 | 102.2486 | 4:09:37 |
| 2.1969 | 102.2486 | 4:09:42 |
| 2.19684 | 102.2485 | 4:09:45 |
| 2.19681 | 102.2485 | 4:09:47 |
| 2.19673 | 102.2485 | 4:09:52 |
| 2.19663 | 102.2484 | 4:09:57 |
| 2.19656 | 102.2483 | 4:10:02 |
| 2.19650 | 102.2483 | 4:10:07 |
| 2.19642 | 102.2482 | 4:10:12 |
| 2.19635 | 102.2482 | 4:10:15 |
| 2.19628 | 102.2481 | 4:10:17 |
| 2.19621 | 102.2480 | 4:10:22 |
| 2.19613 | 102.2480 | 4:10:27 |
| 2.19606 | 102.2479 | 4:10:32 |
| 2.19598 | 102.2479 | 4:10:37 |
| 2.1959 | 102.2478 | 4:10:42 |
| 2.19584 | 102.2478 | 4:10:47 |
| 2.19578 | 102.2477 | 4:10:52 |
| 2.19566 | 102.2476 | 4:10:57 |
| 2.19578 | 102.2475 | 4:11:02 |
| 2.19587 | 102.2474 | 4:11:07 |
| 2.19599 | 102.2473 | 4:11:12 |
| 2.19609 | 102.2472 | 4:11:17 |
| 2.19621 | 102.2470 | 4:11:22 |
| 2.19631 | 102.2469 | 4:11:27 |
| 2.19641 | 102.2468 | 4:11:32 |
| 2.19650 | 102.2467 | 4:11:37 |
| 2.19656 | 102.2467 | 4:11:42 |
| 2.19657 | 102.2466 | 4:11:47 |
| 2.19655 | 102.2467 | 4:11:52 |
| 2.19658 | 102.2466 | 4:11:57 |
| 2.19662 | 102.2466 | 4:12:02 |
| 2.19662 | 102.2466 | 4:12:07 |
| 2.19670 | 102.2465 | 4:12:12 |
| 2.19679 | 102.2464 | 4:12:17 |
| 2.19686 | 102.2464 | 4:12:22 |
| 2.19679 | 102.2463 | 4:12:27 |
| 2.19671 | 102.2462 | 4:12:32 |
| 2.19665 | 102.2462 | 4:12:37 |
| 2.19664 | 102.2462 | 4:12:42 |
| 2.19661 | 102.2461 | 4:12:47 |
| 2.19658 | 102.2461 | 4:12:52 |
| 2.19642 | 102.2460 | 4:12:57 |
| 2.19633 | 102.2459 | 4:13:02 |
| 2.19630 | 102.2459 | 4:13:07 |
| 2.19635 | 102.2458 | 4:13:12 |
| 2.19655 | 102.2458 | 4:13:17 |
| 2.19673 | 102.2458 | 4:13:22 |
| 2.19674 | 102.2459 | 4:13:27 |
| 2.19683 | 102.2460 | 4:13:32 |
| 2.19693 | 102.2461 | 4:13:37 |
| 2.19701 | 102.2461 | 4:13:42 |
| 2.19710 | 102.2461 | 4:13:47 |
| 2.19721 | 102.2460 | 4:13:52 |
| 2.19729 | 102.2459 | 4:13:57 |
| 2.19739 | 102.2459 | 4:14:02 |
| 2.19748 | 102.2458 | 4:14:07 |
| 2.19758 | 102.2457 | 4:14:12 |
| 2.19772 | 102.2457 | 4:14:17 |
| 2.19781 | 102.2458 | 4:14:22 |
| 2.19784 | 102.2459 | 4:14:27 |
| 2.19787 | 102.2460 | 4:14:32 |
| 2.19798 | 102.2460 | 4:14:37 |
| 2.19810 | 102.2461 | 4:14:42 |
| 2.19833 | 102.2461 | 4:14:47 |
| 2.19855 | 102.2462 | 4:14:52 |
| 2.19875 | 102.2462 | 4:14:57 |
| 2.19897 | 102.2463 | 4:15:02 |
| 2.19909 | 102.2463 | 4:15:07 |
| 2.19928 | 102.2464 | 4:15:12 |
| 2.19940 | 102.2464 | 4:15:17 |
| 2.19949 | 102.2465 | 4:15:22 |
| 2.19958 | 102.2465 | 4:15:27 |
| 2.19972 | 102.2465 | 4:15:32 |
| 2.19985 | 102.2466 | 4:15:37 |
| 2.19984 | 102.2466 | 4:15:42 |
| 2.19982 | 102.2467 | 4:15:47 |
| 2.19979 | 102.2468 | 4:15:52 |
| 2.19978 | 102.2469 | 4:15:57 |
| 2.19976 | 102.2470 | 4:16:02 |
| 2.19976 | 102.2471 | 4:16:07 |
| 2.19976 | 102.2472 | 4:16:12 |
| 2.19976 | 102.2473 | 4:16:17 |
| 2.19976 | 102.2474 | 4:16:22 |
| 2.19979 | 102.2475 | 4:16:27 |
| 2.19979 | 102.2476 | 4:16:32 |
| 2.19977 | 102.2476 | 4:16:37 |
| 2.19976 | 102.2476 | 4:16:42 |
| 2.19975 | 102.2476 | 4:16:47 |
| 2.19972 | 102.2476 | 4:16:52 |
| 2.19964 | 102.2476 | 4:16:57 |
| 2.19954 | 102.2476 | 4:17:02 |
| 2.19954 | 102.2476 | 4:17:07 |
| 2.19944 | 102.2477 | 4:17:12 |
| 2.19936 | 102.2477 | 4:17:17 |
| 2.19929 | 102.2477 | 4:17:22 |
| 2.19923 | 102.2477 | 4:17:27 |
| 2.19916 | 102.2477 | 4:17:32 |
| 2.19910 | 102.2477 | 4:17:37 |
| 2.19906 | 102.2477 | 4:17:42 |
| 2.19904 | 102.2478 | 4:17:47 |

### 6.5 Test Results and Analysis

The raw data from the GPS trajectories is converted into .gpx format. After converted to .gpx format, the file is then analyzed and visualized. The result is shown in figure 6.2:

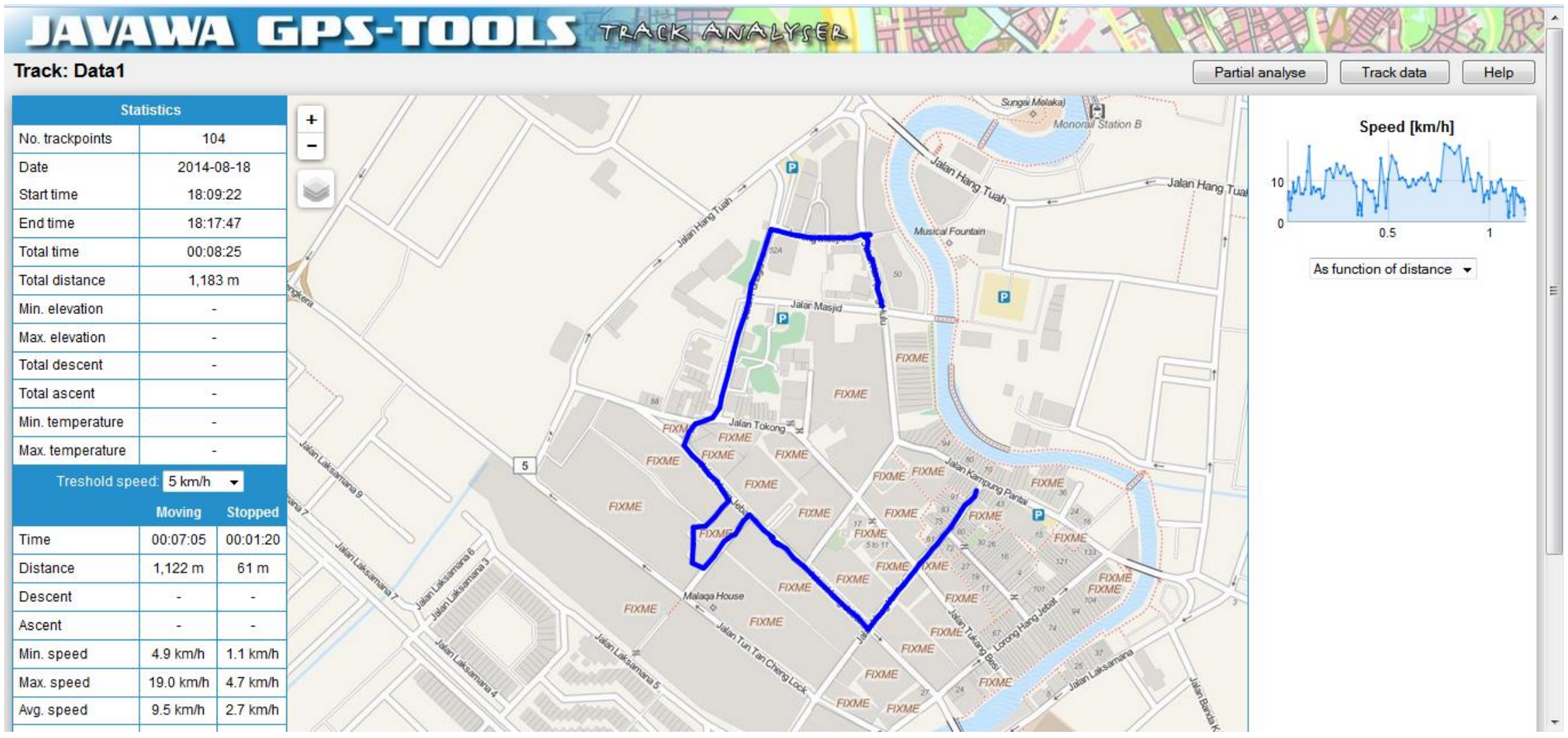

**Figure 6.2: GPS trajectories**

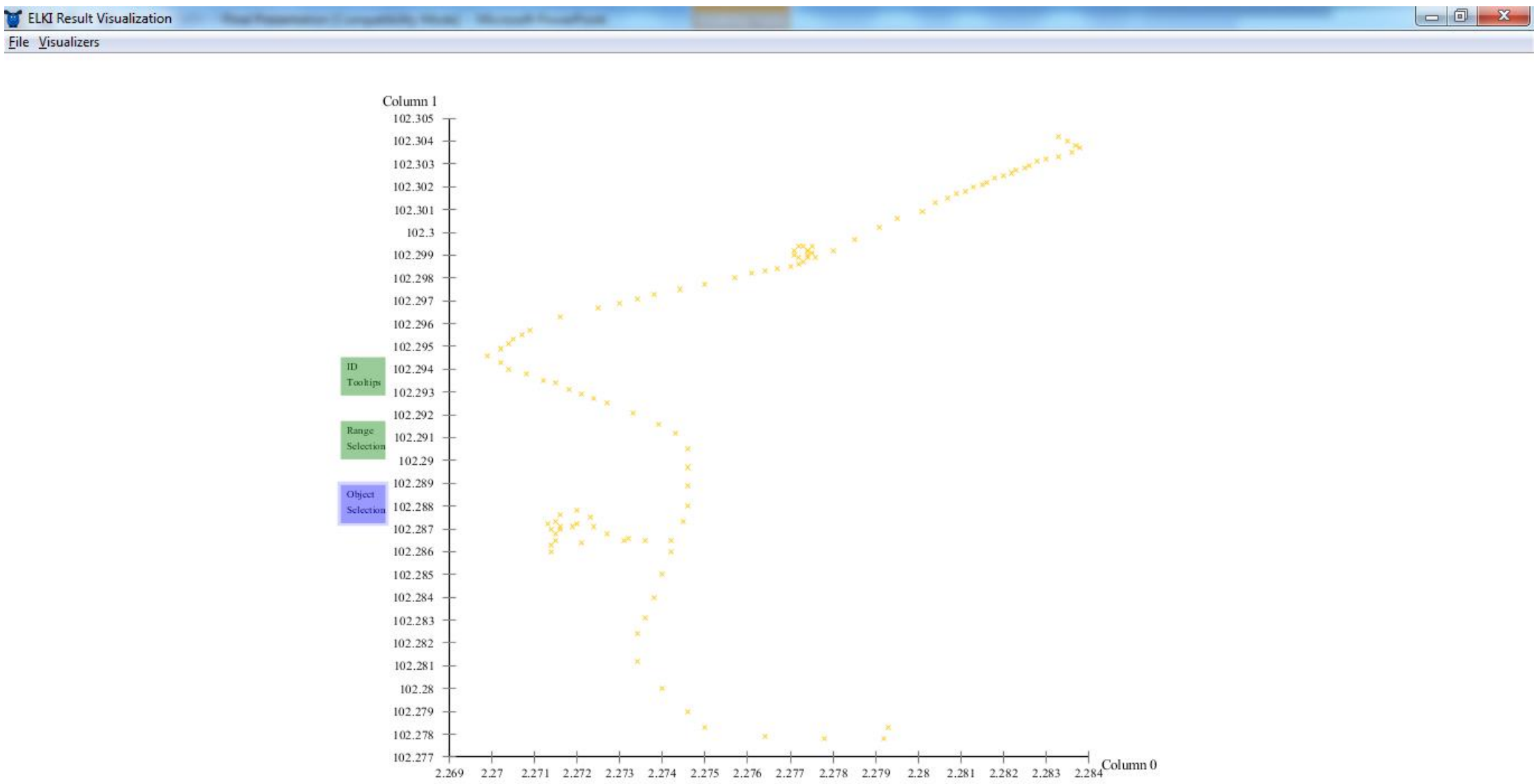

**Figure 6.3: Density based Clustering**

From the figure 6.2, the user GPS trajectory is visualized in figure. The speed of the user that used to travel that area is around 20km/h. In the next figure, density based clustering is used to get the stay point of the user which the user like to travel. From the figure 6.3, can conclude that points of interest are at Planetarium Melaka (2.272, 102.287) and Zoo Melaka (2.2774, 102.299).

**Table 6.5: Result**

| Individual | Number of Paths | Number of Coordinates | Coordinates Selected | Stay Points Detected |
|---|---|---|---|---|
| 1 (000) | 385 | 1010 | 881 | 38 |
| 2 (001) | 859 | 1502 | 795 | 11 |
| 3 (002) | 815 | 1503 | 773 | 27 |
| 4 (003) | 518 | 1232 | 980 | 90 |
| 5 (004) | 937 | 1372 | 1118 | 57 |
| 6 (005) | 901 | 1009 | 873 | 15 |
| 7 (006) | 830 | 1083 | 866 | 31 |
| 8 (007) | 903 | 1903 | 572 | 66 |
| 9 (008) | 899 | 1099 | 695 | 12 |
| 10 (009) | 1006 | 1086 | 947 | 10 |
| Total | 8053 | 12799 | 8500 | 357 |

From the table 6.5, shows that the total paths from 10 users are 8053 and the coordinates detected are 12799. The coordinate selected after pre-processing are 8500. There are 357 stay point detected from total 10 users GPS trajectories. From the 357 stay points, OPTICS clustering is done and gets 14 point of interest. The 14 points of interest are the coordinate point with most density of people that travel to that coordinate which indicate a popular travel location.

## 6.6 Conclusion

This chapter discussed about testing and experiment result. In this chapter, testing phase including test plan, test strategy and test implementation is discussed. At the end of the testing phase, user acceptance test is carrying out.

The first section of this chapter is about test plan. All the plans regarding testing phase is documented. Test plan is designed to perform test. The flow of testing phase is to ensure all the module in application is tested systematically and correctly.

Next section is the test strategy. In this section the strategy used by developer to perform test is recorded as test case. The techniques that are used in the test case are bottom up approach, white box testing, and black box testing.

After that, test data and data analysis is discussed. The GPS trajectories data is process and the result is tested. The analysis of the data is carrying out and documented.

The test result of all test performed is recorded. If the result is fail, the module need to be edit and modified. After that, the test case is carrying out again to ensure there is no major error. After fixed the error and the test case is run smoothly, user acceptance testing is carrying out which involve 10 users.  Improvement is done by getting user comments.

After analysis and test data, the point of interest of 10 users is recorded. Density based clustering method is used to get the point of interests. The coordinate of point of interest is store in the database and when user view the application, the map with point of interest is shown. Refer figure 6.4.

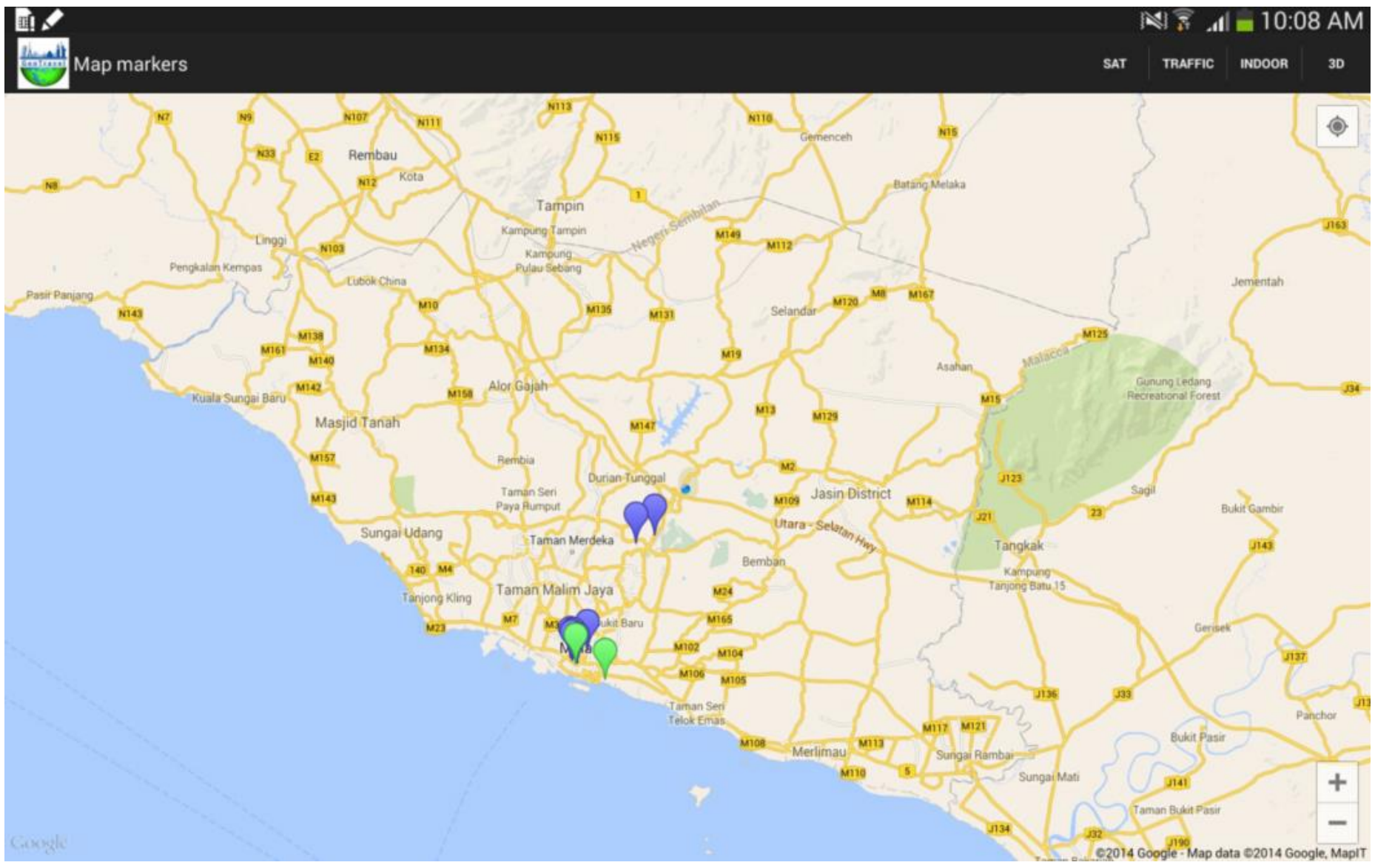

**Figure 6.4: Point of interest**

GeoTravel Application function by auto detected the user location. If the user wishes to travel other place, he can use the search menu to search the location or place he wish to travel. He can enter the place name or the coordinate to search the place. Refer figure to see the detail.

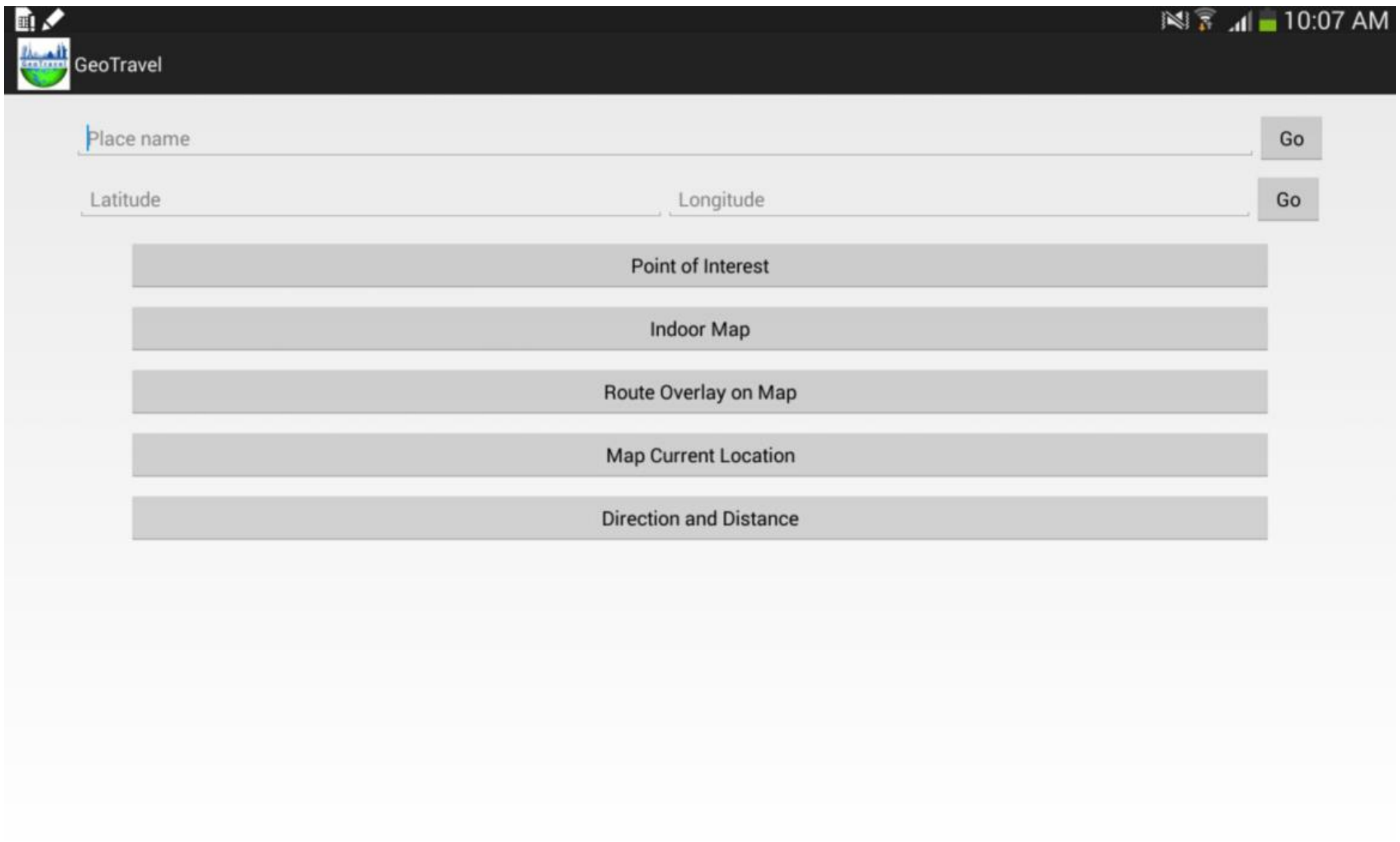

**Figure 6.5: Search Location**

If he wishes to know the point of interest around him, he can click the point of interest button. This module will direct the user to the map and show the point of interest around that area. Refer figure 6.6.

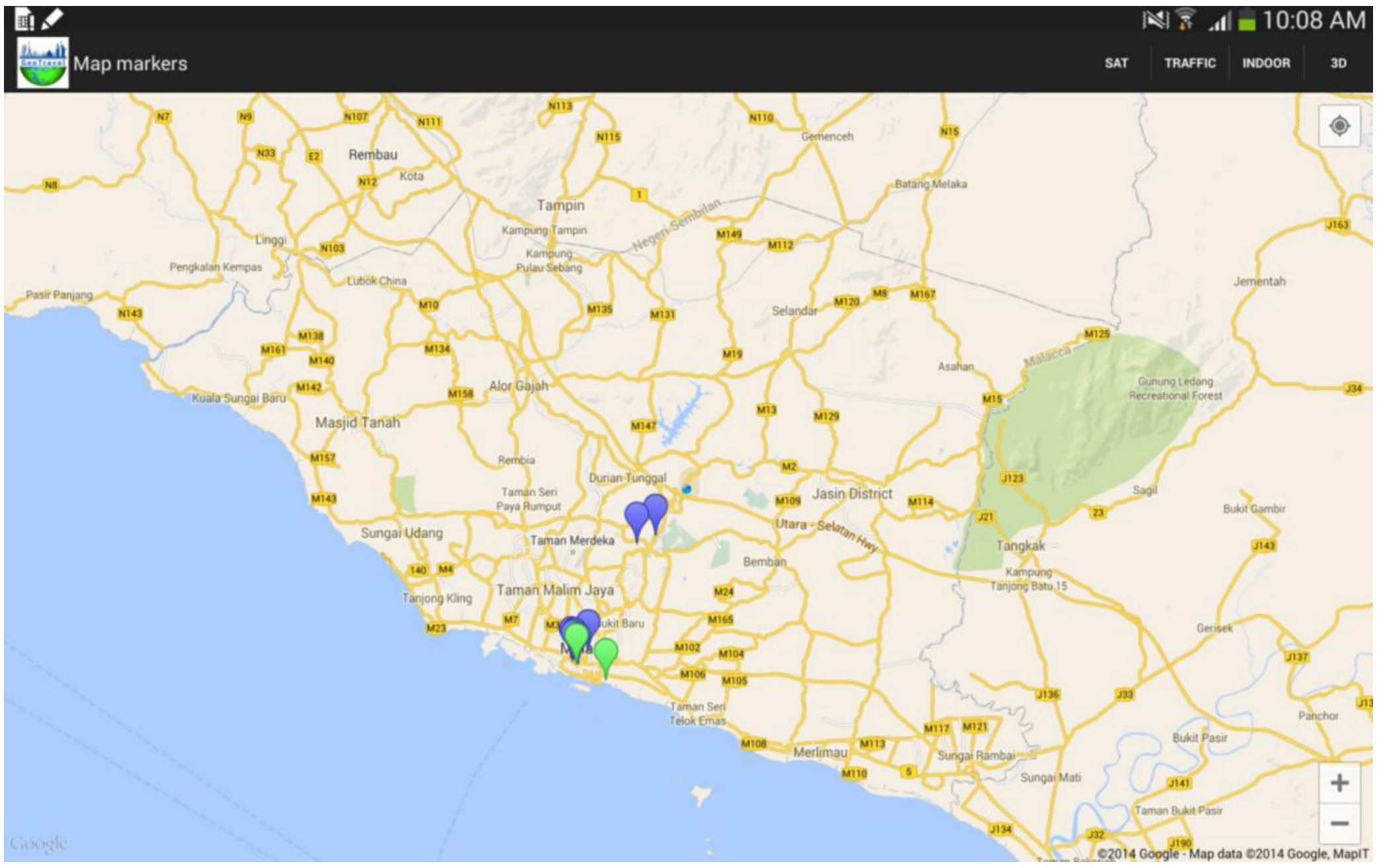

**Figure 6.6: Place of Interest**

After getting the point of interest, if the user wishes to know the traffic allow the point of interest, he can click the traffic button at the upper page. The traffic analysis is show.  Refer figure. The green colour indicates fluent of traffic, and red colour indicates traffic jam.

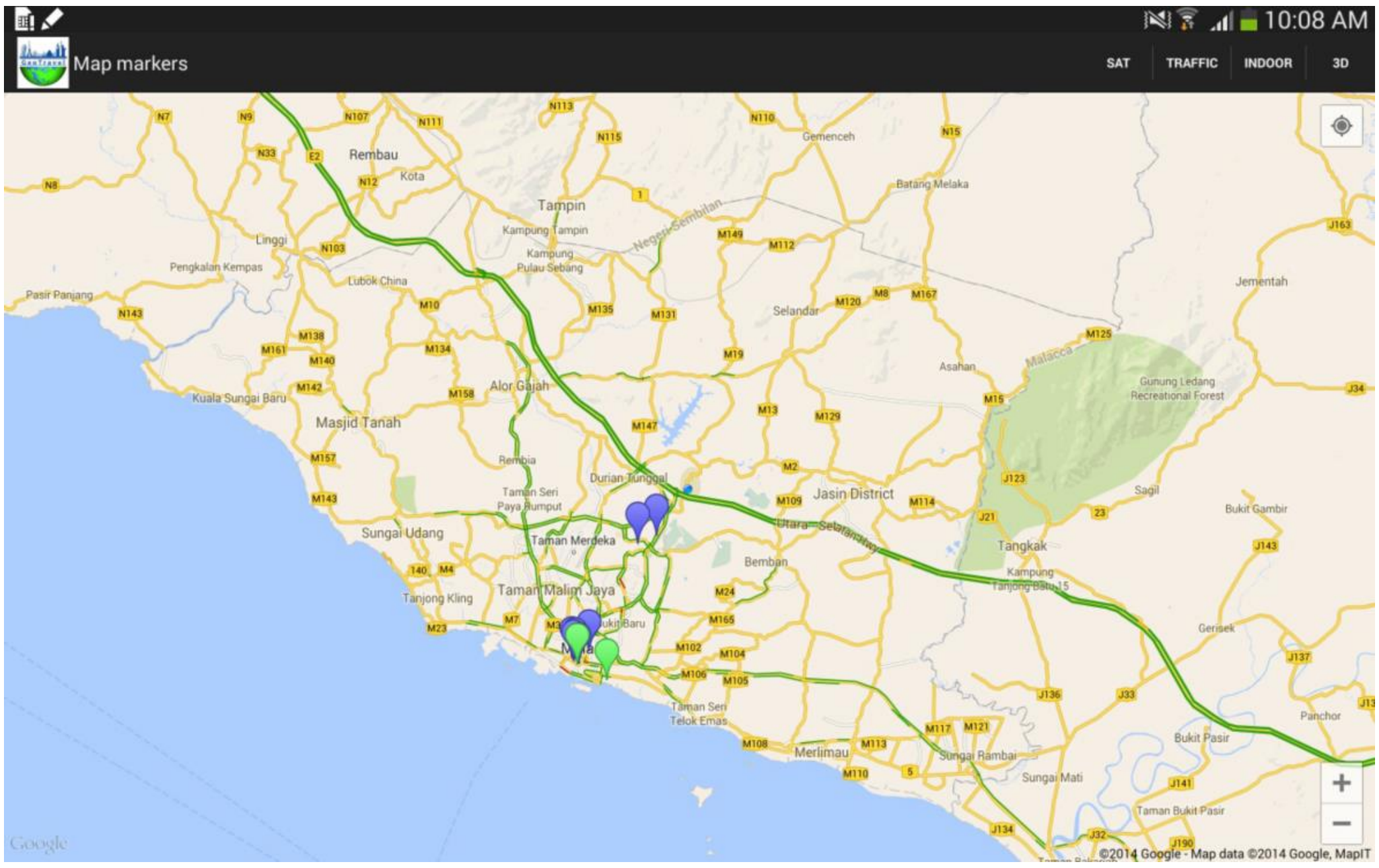

**Figure 6.7: Traffic Analysis**

If the user wants to know the detail about the point of interest, he can click the marker. After the user click, the marker will show the name of place and simple description. Refer figure 6.8. If the user wishes to know more, he can click the box above the marker and the application will direct to the page that describe the point of interest in detail. Refer figure.

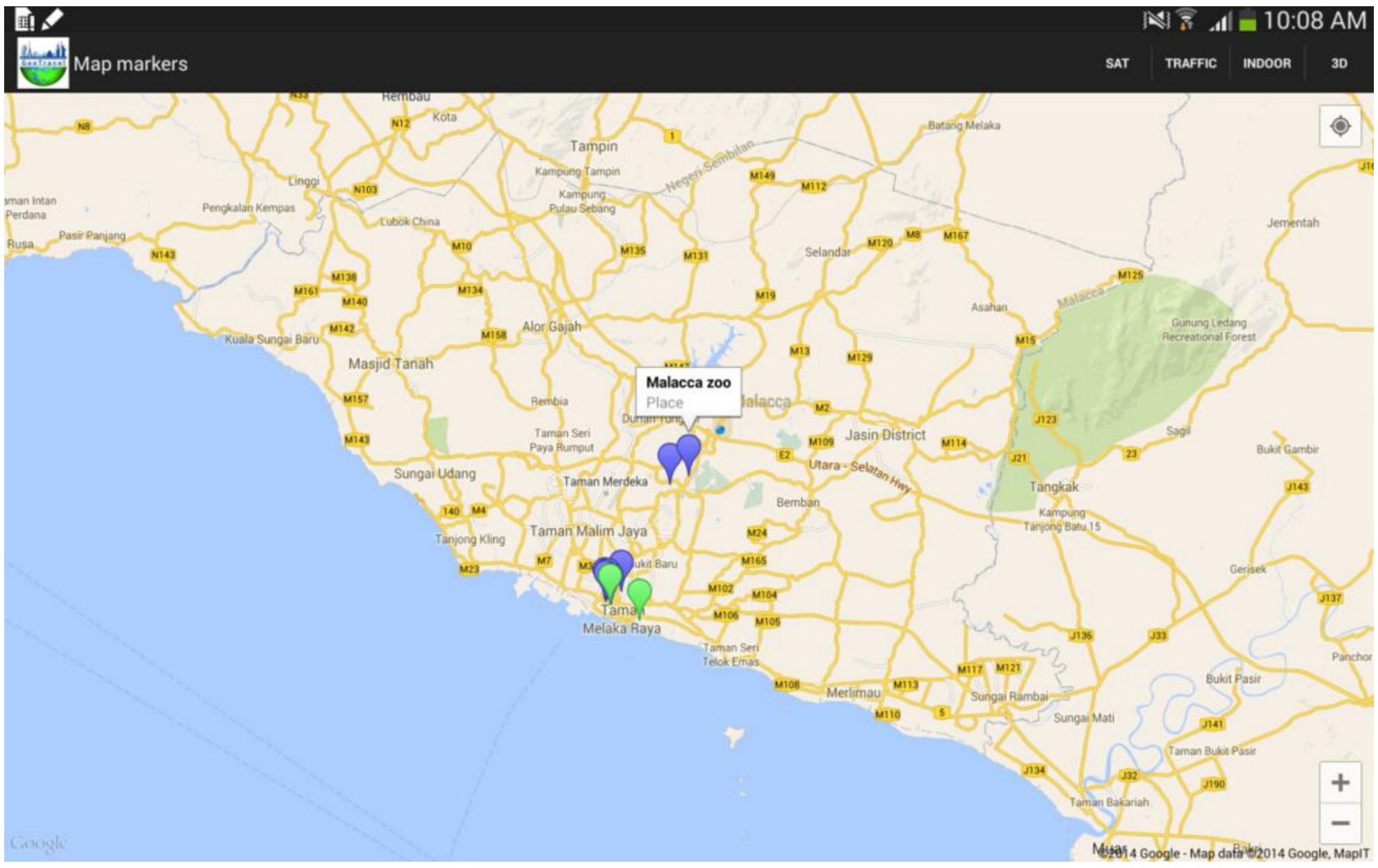

**Figure 6.8: Description of POI**

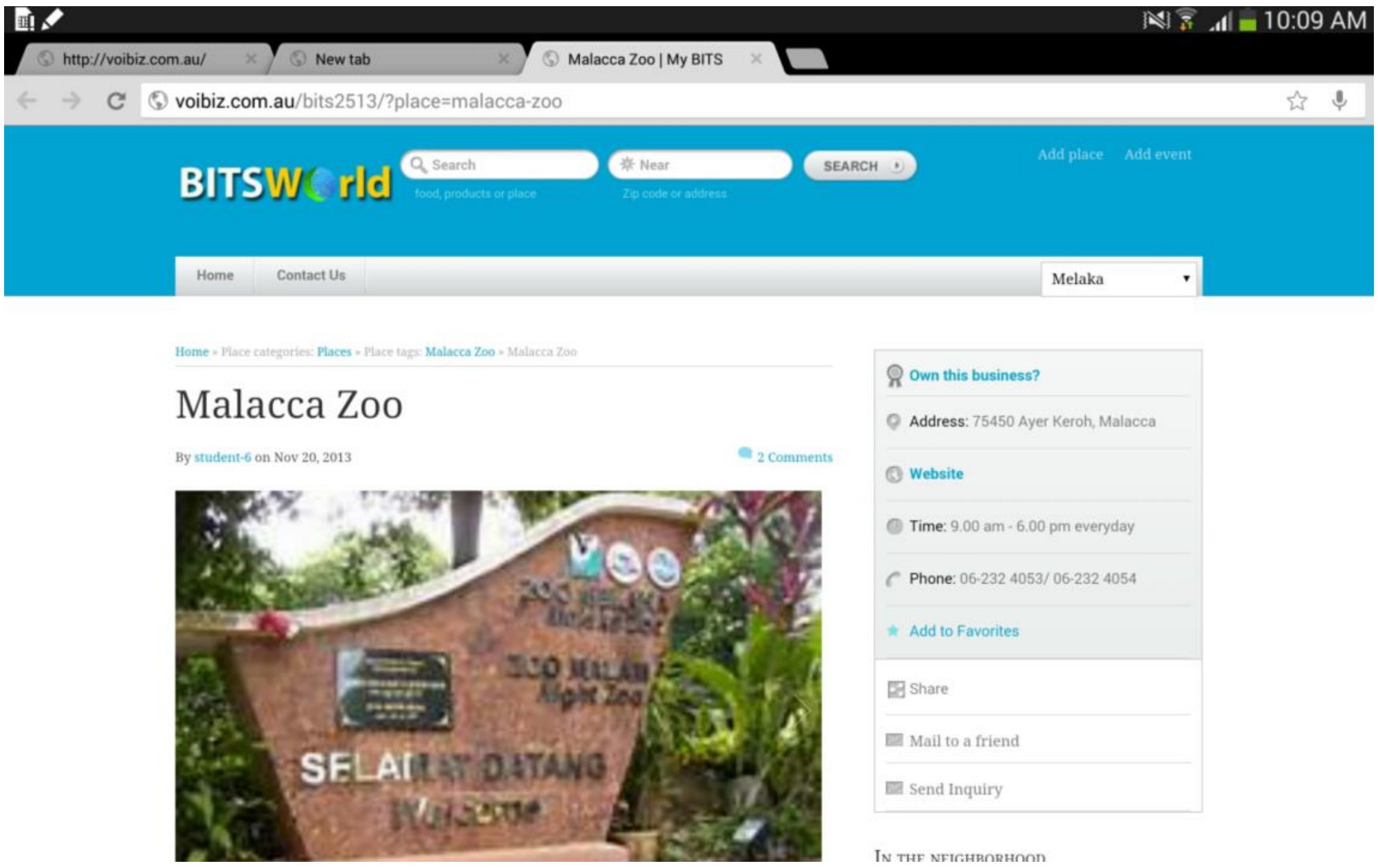

**Figure 6.9: Detail of POI**

The last module of this application is about distance and duration. If the user wishes to know the distance and duration between two points, he can just click two point of interest on the map. The system will calculate the shorter path and show the distance with duration on the screen. This can help the user in plan their trip so that they use their time accurately. Refer figure 6.10.

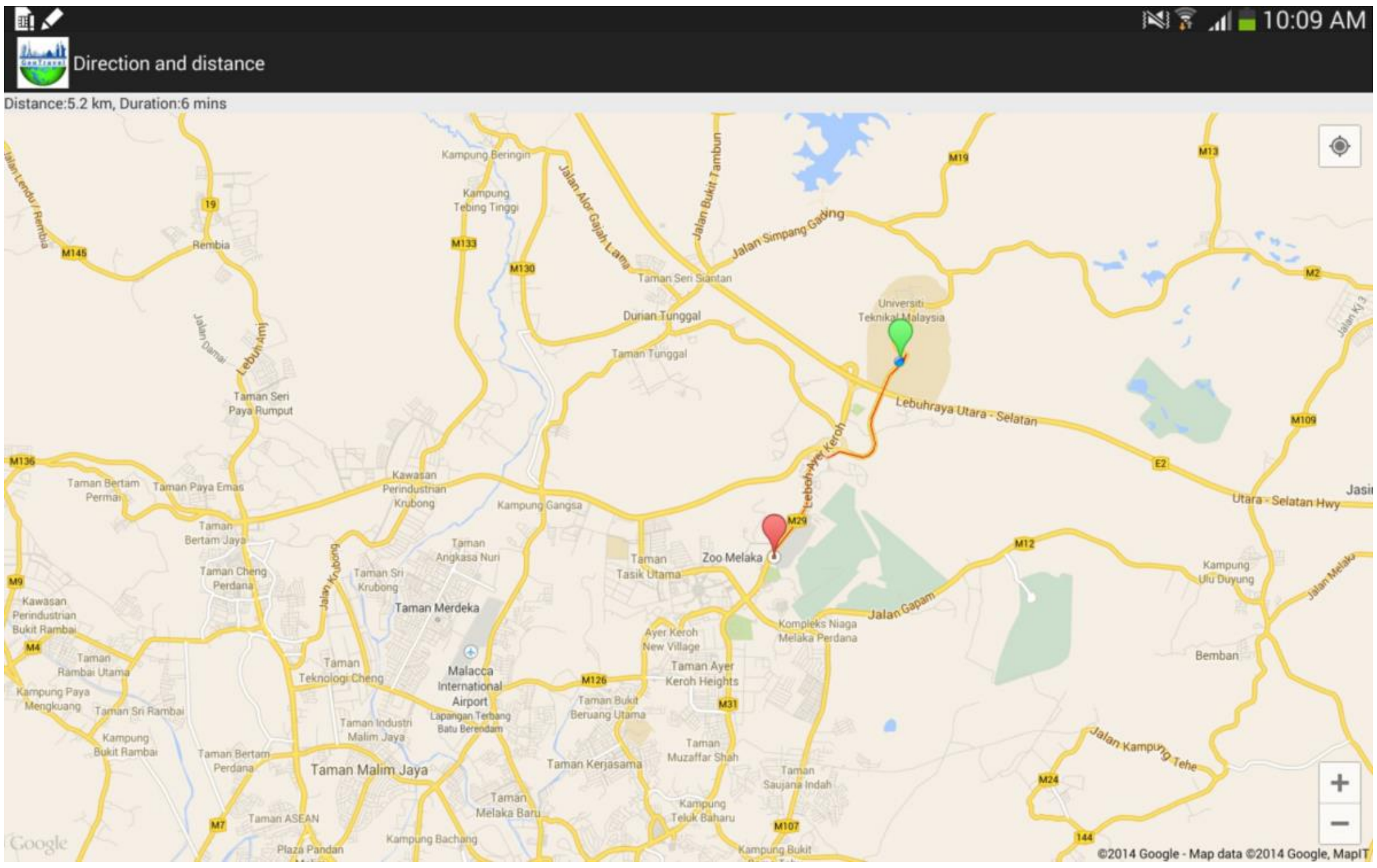

**Figure 6.10: Distance and Duration**

# CHAPTER VII

# PROJECT CONCLUSION

## 7.1 Introduction

Project Conclusion is the last chapter of this project. In this chapter, the conclusion of the whole project is discussed. The strengths and the weaknesses of GeoTravel application is observed and the suggestion of improvement is discussed so that the system can become better.

In the first section of this chapter, strengths and weaknesses of the application are discussed. Every system or application must have strength and weakness. To have a better application, developer must maintain the strength and improve the weakness. The strength and weakness include in architecture, algorithm, user interface and database. Limitation of GeoTravel application will be discussed too.

Next section is about improvement of the system based on the limitation. This section discussed the improvement that can be done to enhance the system. Contribution of the system that can be made is also discussed in this chapter. The contribution of GPS trajectories and GeoTravel application is recorded.

## 7.1 Observation on Weaknesses and Strengths

### 7.1.1 Strengths in GeoTravel System

Strength of GeoTravel System includes it consists of web-based and android application. This feature gives another choice for the user. The android application is light compare to web-based, therefore it is easier to use.

Besides that, point of interest of GeoTravel System is based on mining of GPS trajectories. It also included the live event which many systems does not include. The live event that held in a place is important for a travelers because some of the traveler especially backpacker may want to join local event to gain more experience. By mining of GPS trajectories, the point of interest that suggest is the popular place. This can give an idea to people who are not familiar with the places.

Other than this, GeoTravel System also included a multiday itineraries plan. User can choose place to visit based on their preferences and the system will plan the itineraries for them.

### 7.1.2 Weakness in GeoTravel System

The weakness in GeoTravel System is there is current application was developed for state of Melaka. This is because data collections of real GPS trajectory take a lot of time. To obtain more point of interest, it requires more GPS trajectory.

Besides that, GeoTravel application required internet connection to work. This included the web-based or android version. This is because checking user location and getting information from the map required internet connection. Without internet connection, most of the task cannot be done.

The GPS trajectories data are taken from Garmin GPS device. The limitation of this is the data that use to suggest the point of interest is just based on Garmin GPS user. Data from other GPS device is not used because there is only Garmin provided GPS trajectories and this may causes the data only depend on Garmin GPS device.

Other than this, this application do not recorded user trajectory. By record user trajectory, the point of interest may update always based on user travel sequence so that the data is up to date.

## 7.2 Propositions for Improvement

There are strengths and weaknesses in all of the application. If there are weaknesses, then improvement must be made so that the application becomes better. The improvements that can be done on GeoTravel application are:

1. Improve the system so that it support more state or country
2. Improve the code structure and algorithm of the sytem
3. Collect data from more GPS device
4. Enhance the application so that it record user trajectory
5. Develop application that supported other than Android

## 7.3 Contribution

GeoTravel System acts as a platform that provides convenience for traveler and backer. It suggests point of interests based on the GPS trajectories. GeoTravel systems not only contain web system, it also contains android application. This allows user to access the system anytime and anywhere when there is internet connection.

GPS trajectories from various users contain a lot of information. From the information, it helps users to understand the surrounding location and enable travel recommendation. Backpacker needs an efficient and economic trip plan application.

The live event is included so that travellers have more choice. GeoTravel system also included multi-day itineraries plan.

## 7.4 Conclusion

In this chapter, strengths and weaknesses of GeoTravel system is discussed. Strengths of GeoTravel application includes it consists of web-based and android application that suggest point of interest based on GPS trajectories. This feature gives another choice for the user. The android application is light compare to web-based, therefore it is easier to use.

In this chapter, improvement which can be done based on the weaknesses are suggested and discussed. The improvement on server and client side is suggested. This application should support platform other than Android in the future.

In conclusion, this application is better than other travel application is that the point of interest is suggested based on other user GPS trajectories. This mean that only the popular point of interest which most of the users visiting are show and suggested to user.

# APPENDIX 1

Flow Chart and Grantt Chart

# Appendix

## Flow Chart of Project Activities

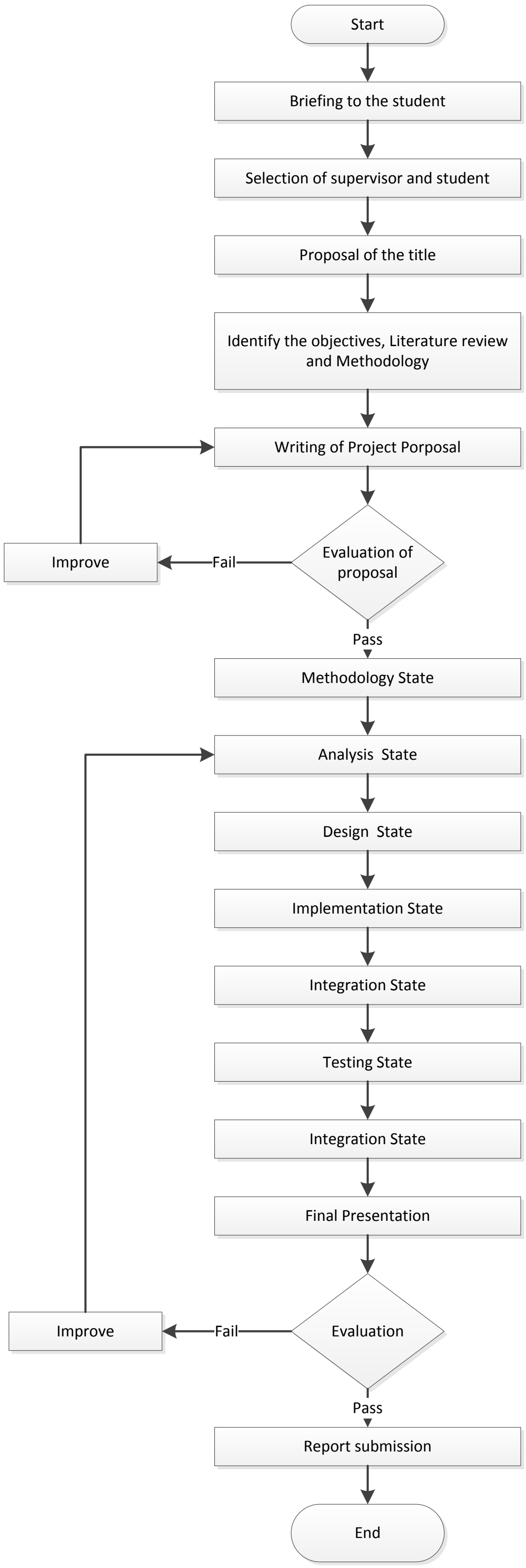

**Gantt Chart of Project Activities**

| Week / Task | 1 | 2 | 3 | 4 | 5 | 6 | 7 | 8 | 9 | 10 | 11 | 12 | 13 | 14 |
|---|---|---|---|---|---|---|---|---|---|---|---|---|---|---|
| **PSM Proposal** | ■ | | | | | | | | | | | | | |
| **Introduction** | | ■ | | | | | | | | | | | | |
| **Methodology stage** | | | ■ | ■ | | | | | | | | | | |
| **Analysis stage** | | | | | ■ | ■ | | | | | | | | |
| **Design stage** | | | | | | | ■ | | | | | | | |
| **Implementation stage** | | | | | | | | ■ | ■ | | | | | |
| **Integration stage** | | | | | | | | | | ■ | | | | |
| **Testing stage** | | | | | | | | | | | ■ | | | |
| **System Demonstration** | | | | | | | | | | | | ■ | | |
| **Presentation** | | | | | | | | | | | | | ■ | |
| **Final report** | | | | | | | | | | | | | | ■ |

# APPENDIX 2

USER MANUAL

# USER MANUAL

Follow the below steps to use the application. GeoTravel application does not required user to login.

1. Choose the GeoTravel application in your device.

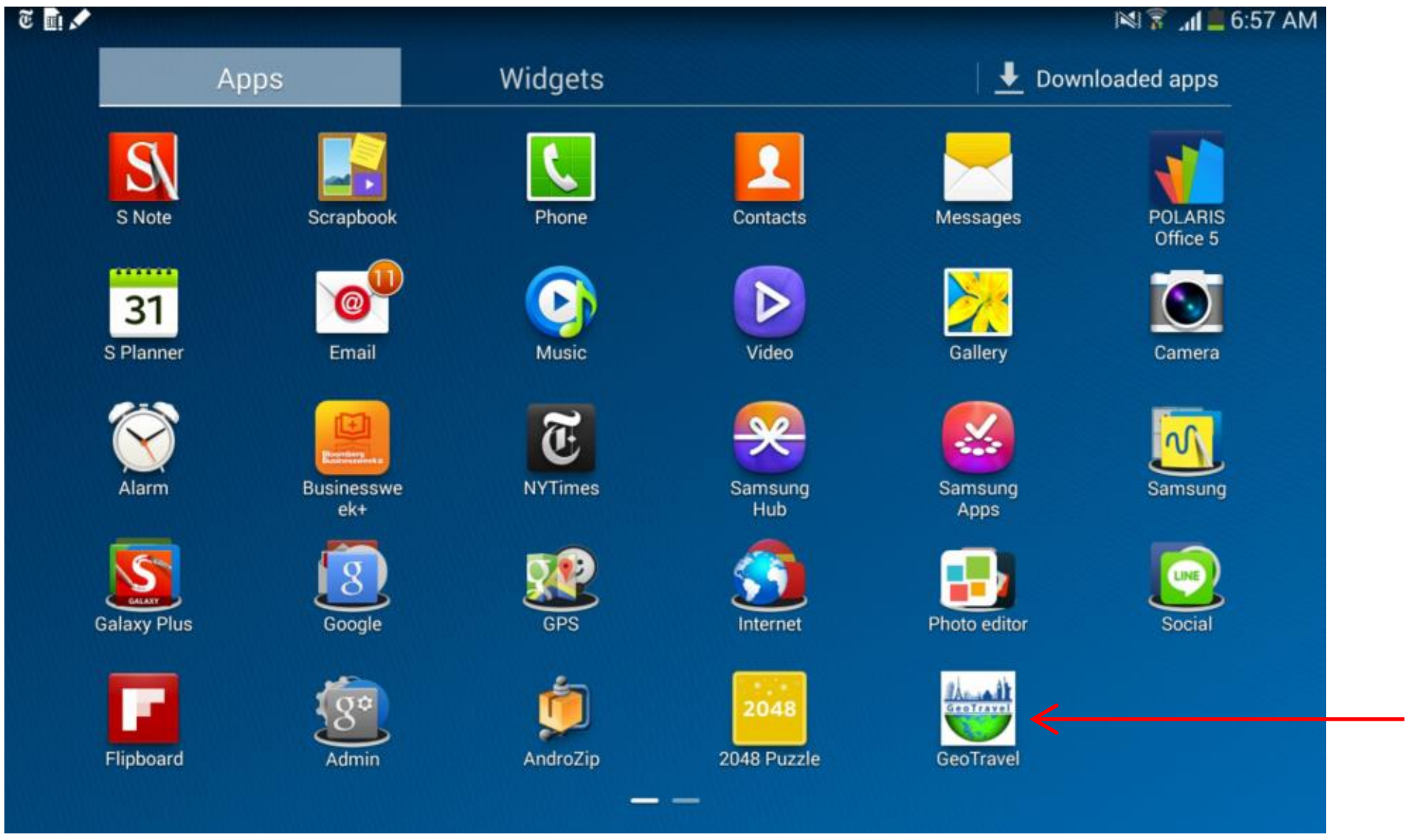

2. Click the welcome button at the first page.

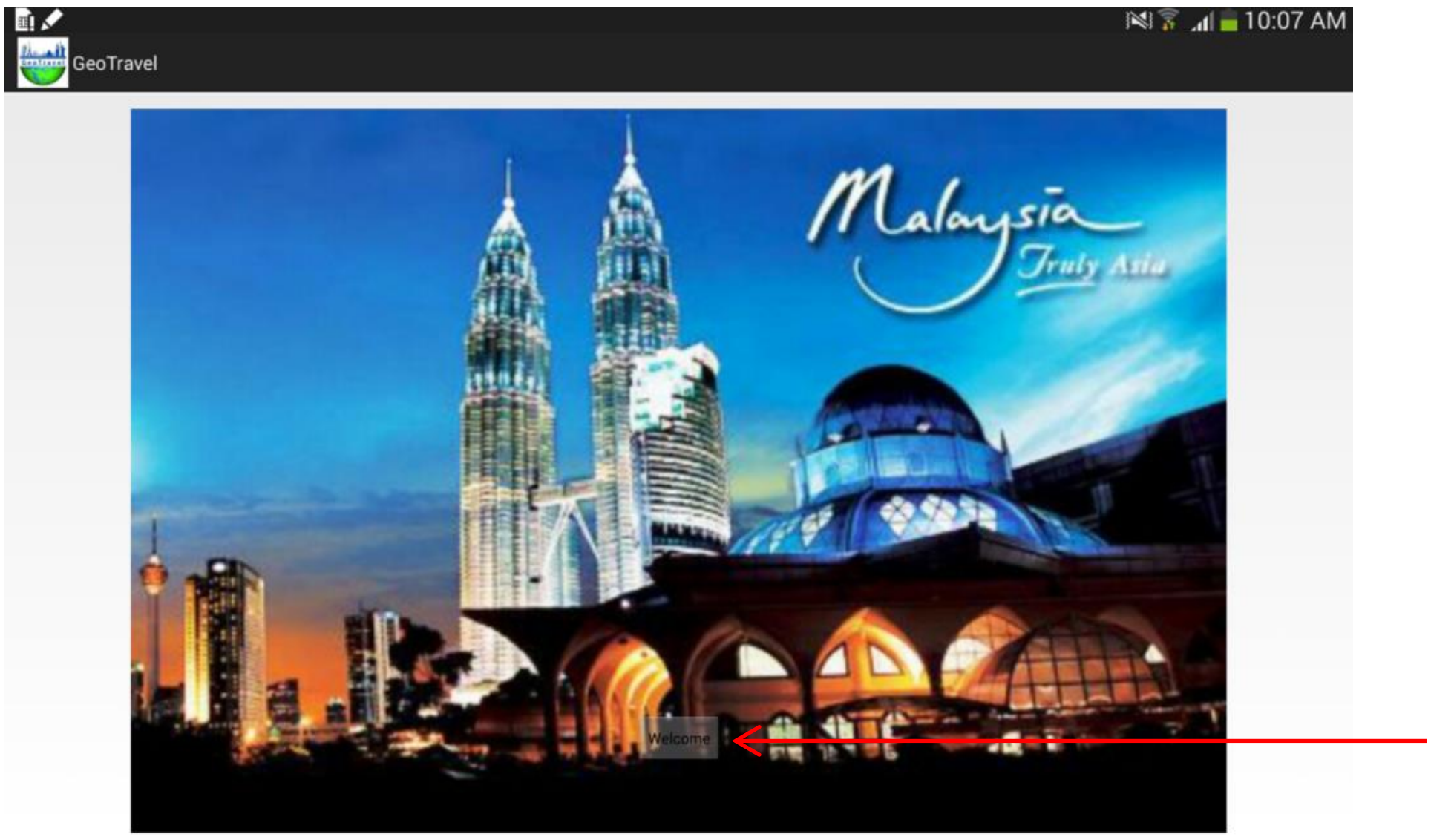

3. User can choose to search the location by enter name of the place.

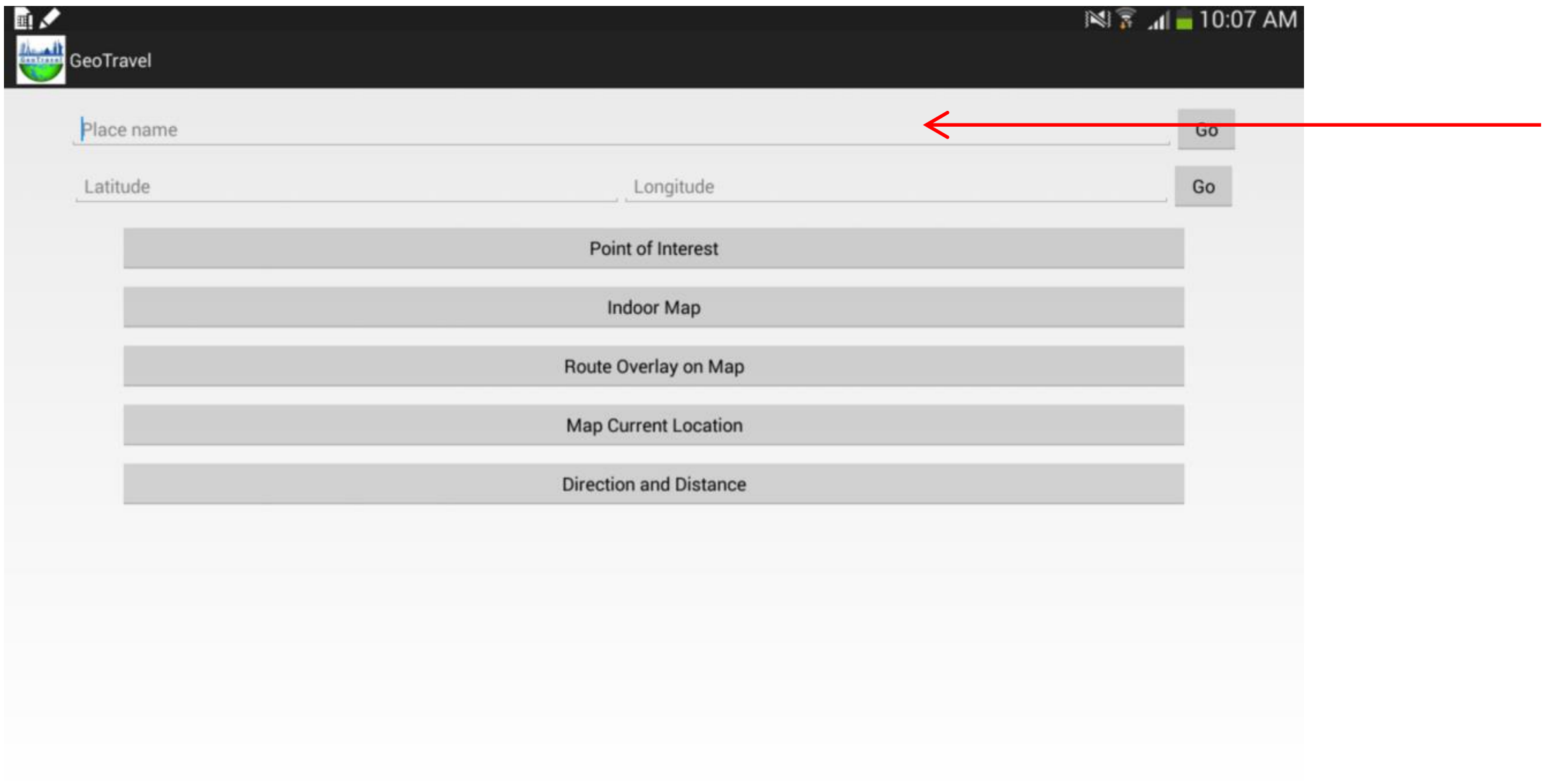

4. User can also see the suggested point of interest that around the user by choose the button Point of interest

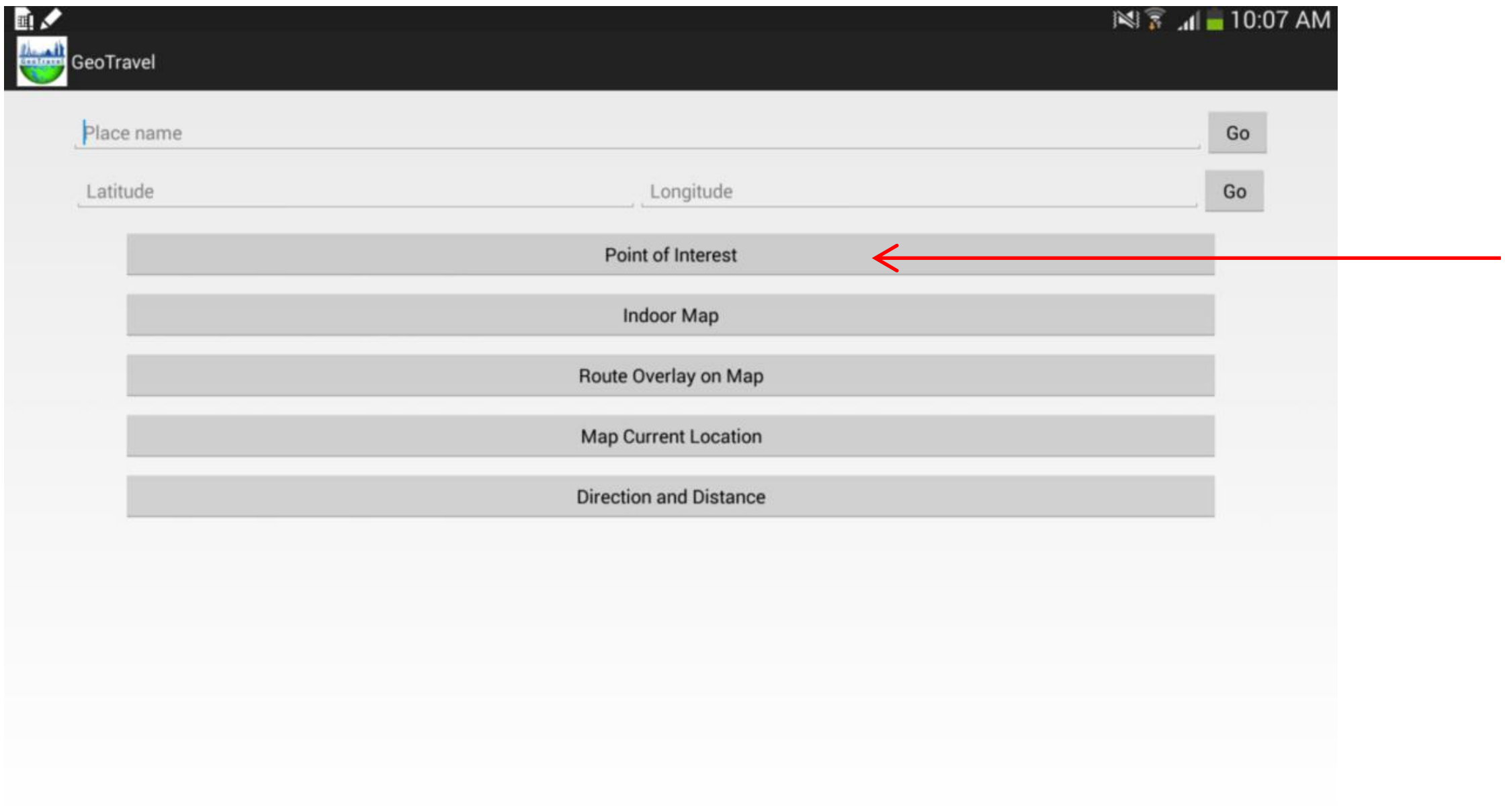

5. The point of interest is show in the map.

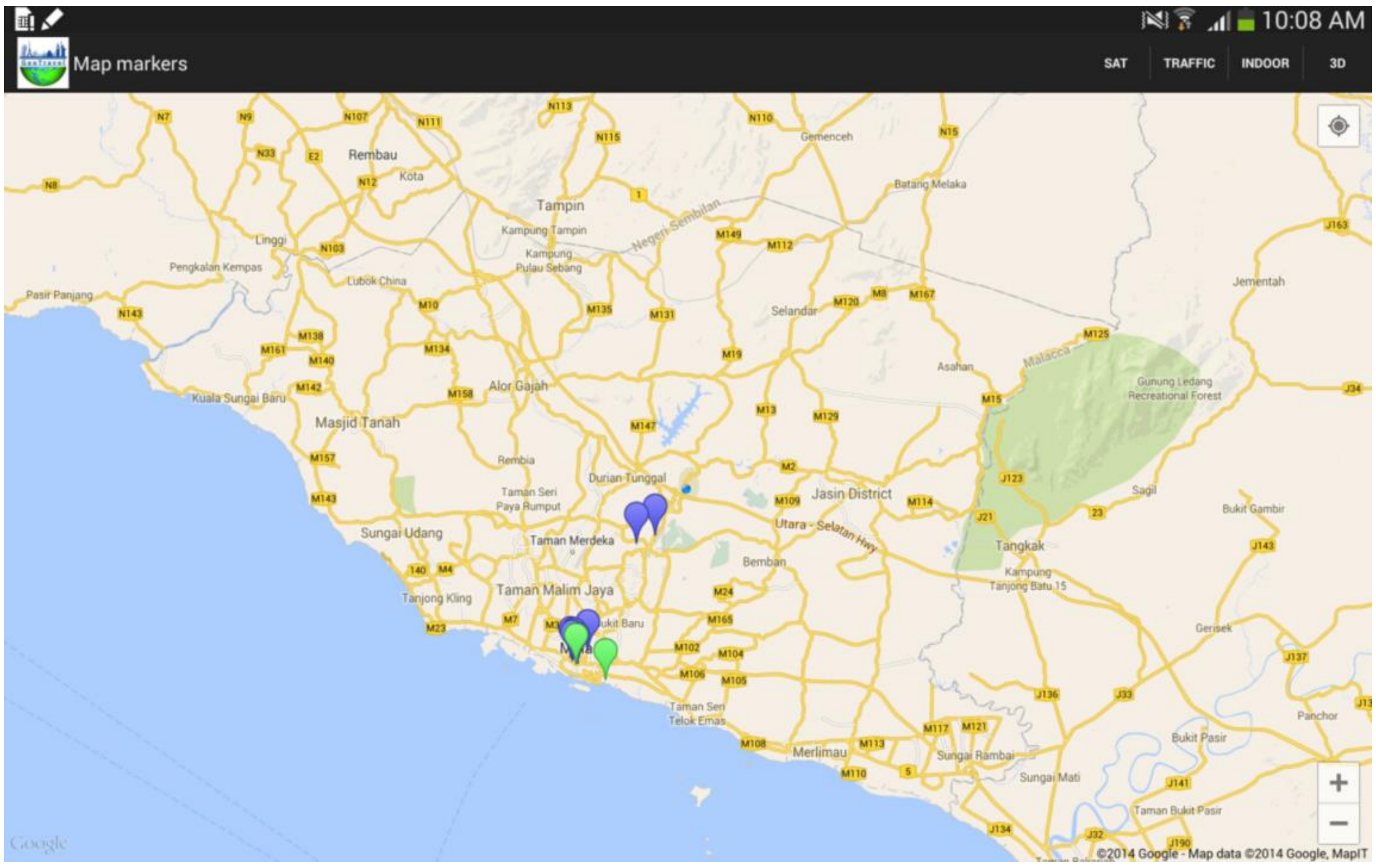

6. User can view the traffic by click the button traffic at top.

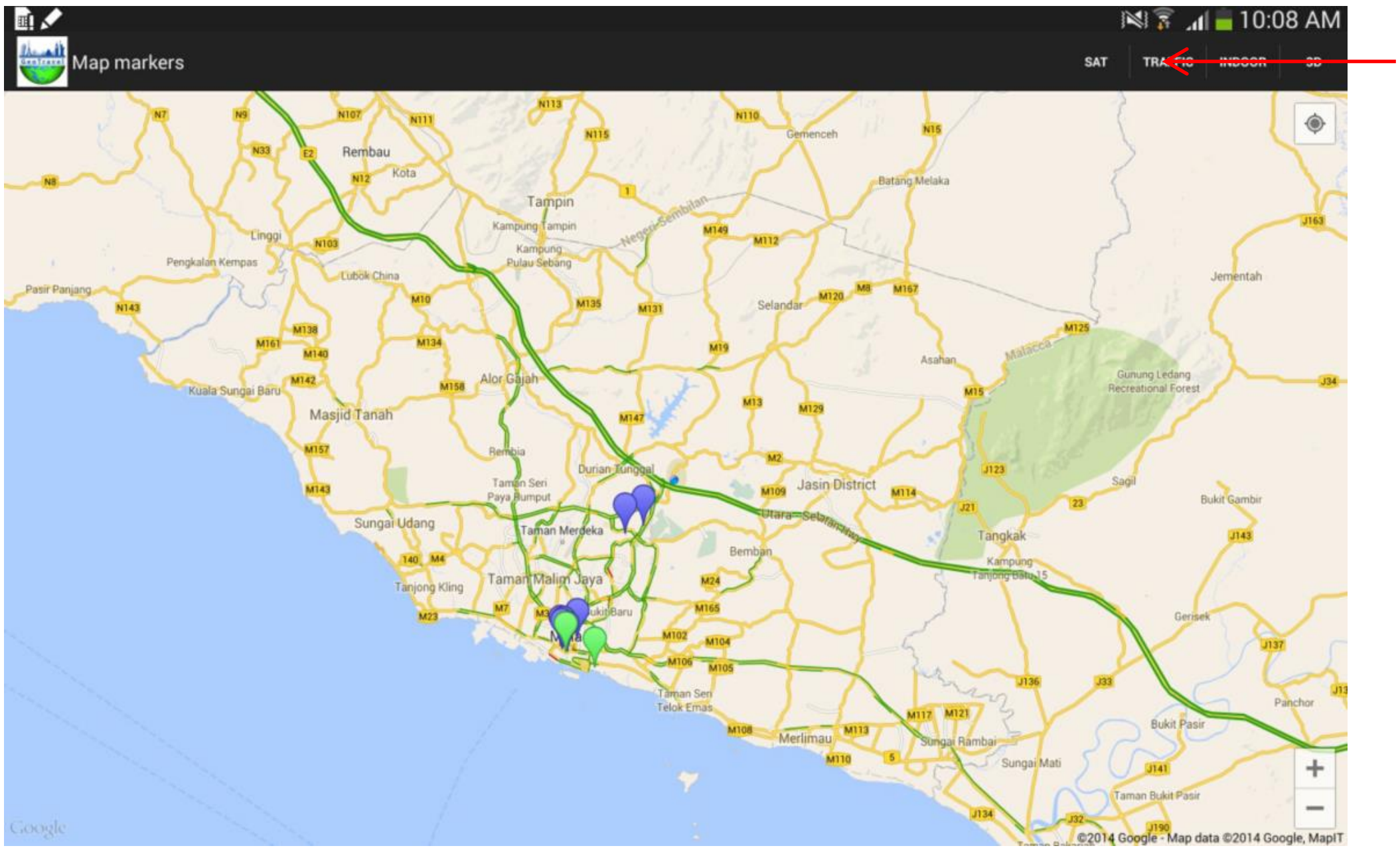

7. User can view the detail by click the marker.

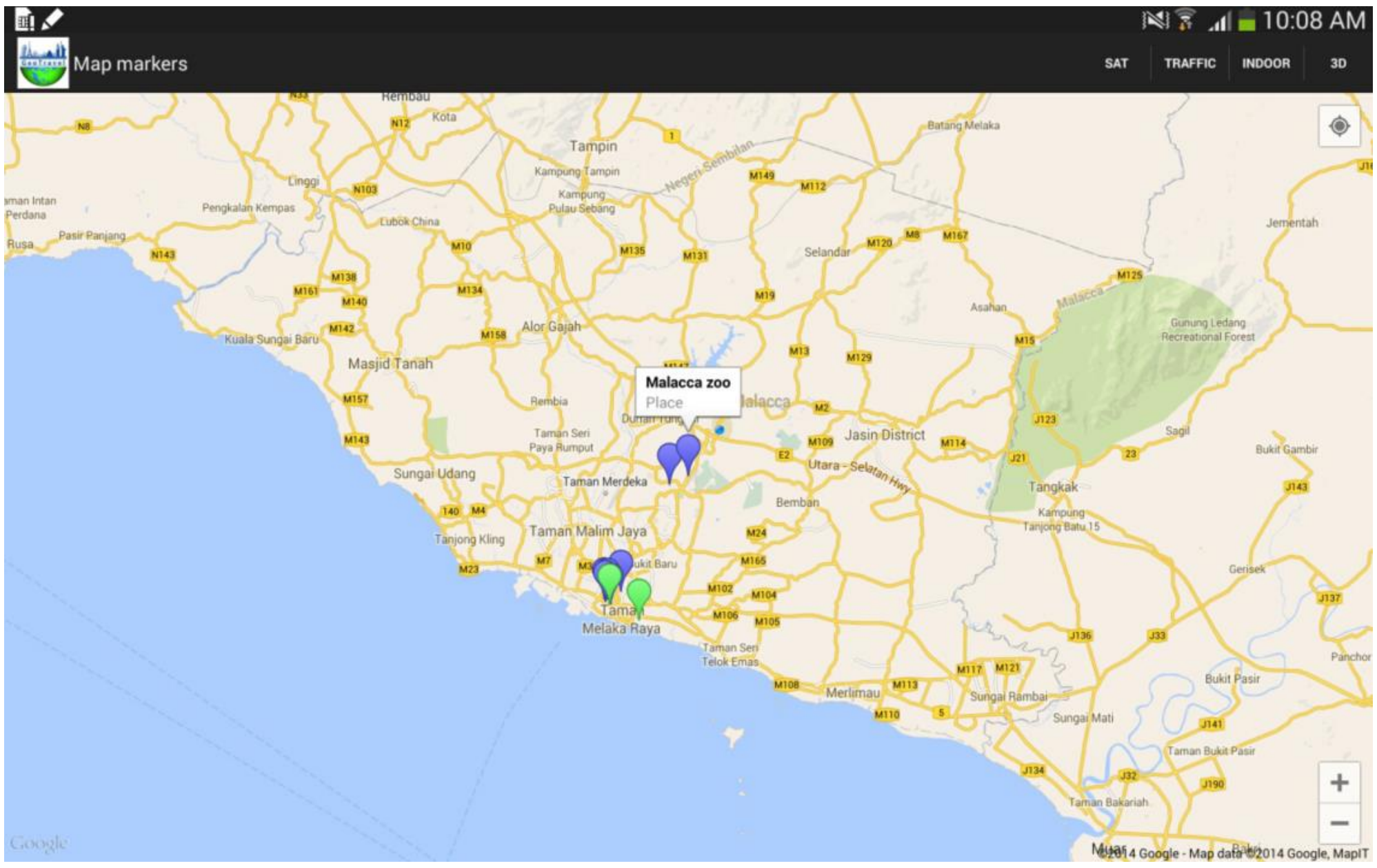

8. Direction and Distance is show when user clicks the button at menu.

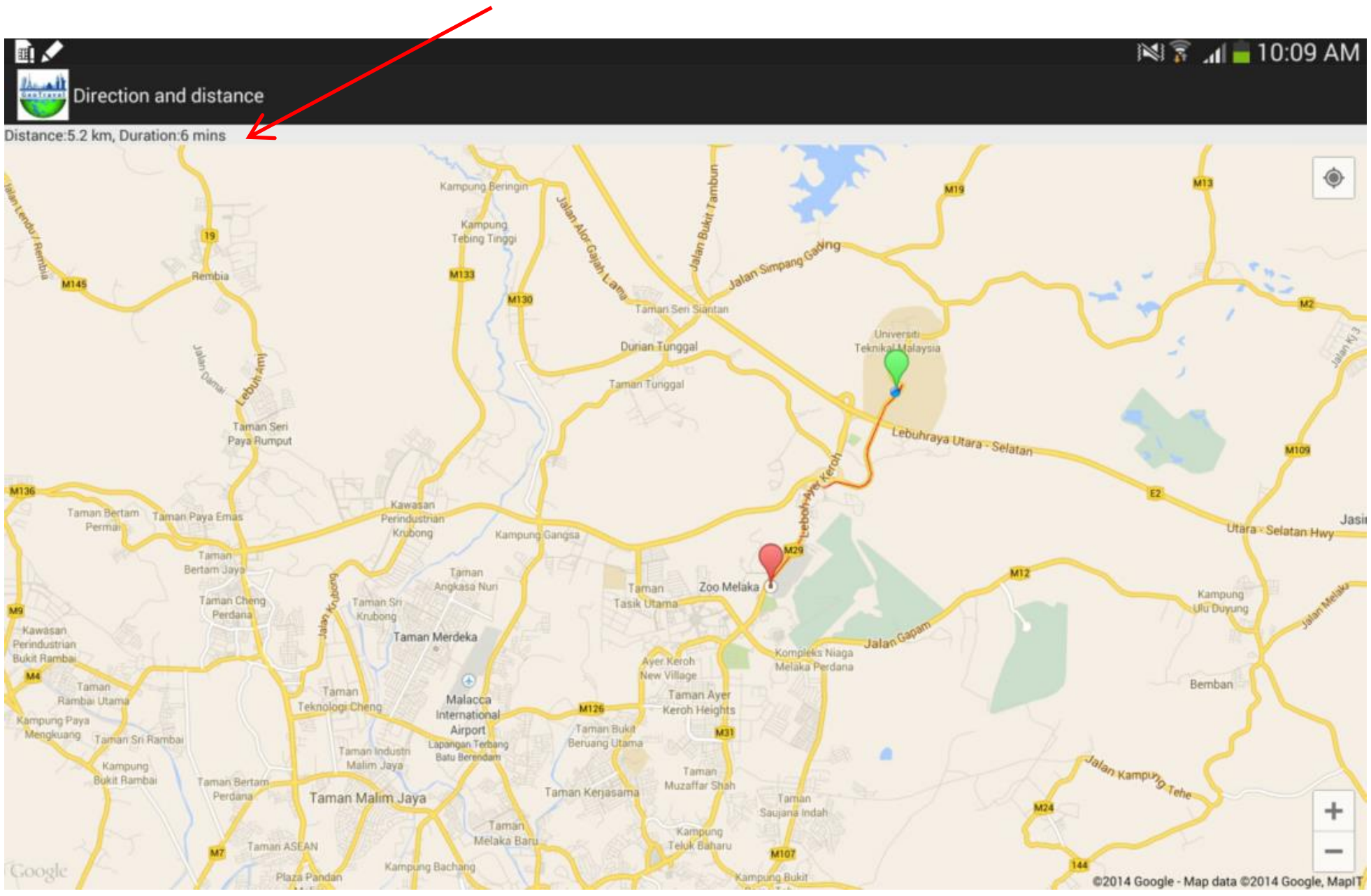